\documentclass[prb,twocolumn,showpacs,superscriptaddress,floatfix]{revtex4-1}
\usepackage{graphicx,amsfonts,amssymb,amsmath,xspace,dsfont}
\usepackage[colorlinks=true,citecolor=blue,linkcolor=red,urlcolor=blue]{hyperref}
\usepackage{color}

\newlength{\ldag}
\begin{document}

\title{Exotic magnetisation plateaus in a quasi-2D Shastry-Sutherland model}

\author{G.R. Foltin}
\affiliation{Lehrstuhl f\"ur Theoretische Physik 1, TU Dortmund, 44221 Dortmund, Germany}
\author{S.R. Manmana}
\affiliation{Institut f\"ur Theoretische Physik, Georg-August-Universit\"at G\"ottingen, 37077 G\"ottingen, Germany}
\author{K.P. Schmidt}
\email{kai.schmidt@tu-dortmund.de}
\affiliation{Lehrstuhl f\"ur Theoretische Physik 1, TU Dortmund, 44221 Dortmund, Germany}

\date{\rm\today}

%------------------------------------------------------------------------------

\begin{abstract}
We find unconventional Mott insulators in a quasi-2D version of the Shastry-Sutherland model in a magnetic field. In our realization on a 4-leg tube geometry, these are stabilized by correlated hopping of localized magnetic excitations. Using perturbative continuous unitary transformations (pCUTs, plus classical approximation or exact diagonalization) and the density matrix renormalisation group method (DMRG), we identify prominent magnetization plateaus at magnetizations $M=1/8$, \mbox{$M=3/16$}, $M=1/4$, and $M=1/2$. While the plateau at $M=1/4$ can be understood in a semi-classical fashion in terms of diagonal stripes, the plateau at $M=1/8$ displays highly entangled wheels in the transverse direction of the tube. Finally, the $M=3/16$ plateau is most likely to be viewed as a classical 1/8 structure on which additional triplets are fully delocalized around the tube. The classical approximation of the effective model fails to describe all these plateau structures which benefit from correlated hopping. We relate our findings to the full 2D system, which is the underlying model for the frustrated quantum magnet SrCu(BO$_3$)$_2$. 
\end{abstract}


\maketitle

%%%%%%%%%%%%%%
\section{Introduction}
%%%%%%%%%%%%%%
%
%

A particularly interesting realization of highly frustrated quantum magnetism\cite{book_quantummagnetism} is found in the correlated material SrCu(BO$_3$)$_2$.\cite{Kageyama99,Onizuka00,Kageyama00,Kodama02,Takigawa04,Levy08,Sebastian08,Jaime12,Takigawa13,Matsuda13} 
The underlying model is widely believed to be the $S=1/2$ Heisenberg antiferromagnet (HAFM) on the 2D Shastry-Sutherland lattice and experiments in ultrastrong magnetic fields unveil a multitude of intriguing behavior archetypical for frustrated quantum magnetism. 
Despite a huge body of literature,\cite{Kageyama99,Onizuka00,Kageyama00,Kodama02,Takigawa04,Levy08,Sebastian08,Jaime12,Takigawa13,Matsuda13,Miyahara99,Momoi00a,Momoi00b,Fukumoto00,Fukumoto01,Miyahara03a,Miyahara03b,Dorier08,Abendschein08,Nemec12,Lou12,Corboz14} still various aspects refrain from a clear theoretical understanding, in particular the structures and sequence of Mott insulators realized as magnetization plateaus in the low part of the magnetization curve, so that the system stays in the focus of present day research.  

The Shastry-Sutherland model \cite{Shastry81} can be seen as a set of mutually orthogonal dimers which are coupled by an interdimer coupling $J'$. 
Its beauty arises from the exact solution in terms of a product state of singlets at zero magnetic field and small enough values of $J'$, and the fact that the magnetization process can be viewed as a subsequent population of the dimers by triplets. 
Even though these aspects give important insights into the magnetization process, the interdimer coupling in the material is so strong that a treatment by analytical methods as well as by numerical approaches has remained a challenge. Here, we combine pCUTs\cite{Knetter00,Knetter03} and DMRG\cite{DMRG} to treat the system in an approximation of the 2D structure. While the DMRG works best in 1D, the uncovering of a spin liquid phase in the HAFM on a 2D kagome lattice\cite{WhiteScienceKagome} shows that this method can lead to insightful results also in higher dimensions, see also Ref.~\onlinecite{Stoudenmire2D}. 
In contrast to this study, here we do not restrict ourselves to the ground state at zero magnetic field but 
address the aforementioned behaviour at finite magnetizations. 
This further degree of freedom makes a systematic study of the full 2D system even more difficult. 
In consequence, we choose to study a quasi-2D version of the Shastry-Sutherland model on a tube geometry which has a finite width of four dimers and is periodically coupled in the transverse direction, and which we refer to as a four-leg tube. 
Note that in the present paper we go one step further than in previous work \cite{Matsuda13} and do not only focus on the magnetization curve itself, but also treat the magnetization structures on the plateaus in high detail and accuracy. 
We believe that this can lead to insights for the full 2D case. 
In particular, we identify a mechanism which leads to {\it delocalized} structures on magnetisation plateaus, and which stabilises highly entangled states in these Mott insulators similar to the recently discovered crystals of bound states\cite{Manmana11} in the 2D Shastry-Sutherland model \cite{Corboz14}. 
The scope of this paper is to discuss this mechanism in detail and propose possible scenarios for the full 2D system.  
 
% 
%
%%%%%%%%%%%%%%
\section{Model}
\begin{figure}[t]
\centering
\includegraphics[width=\columnwidth]{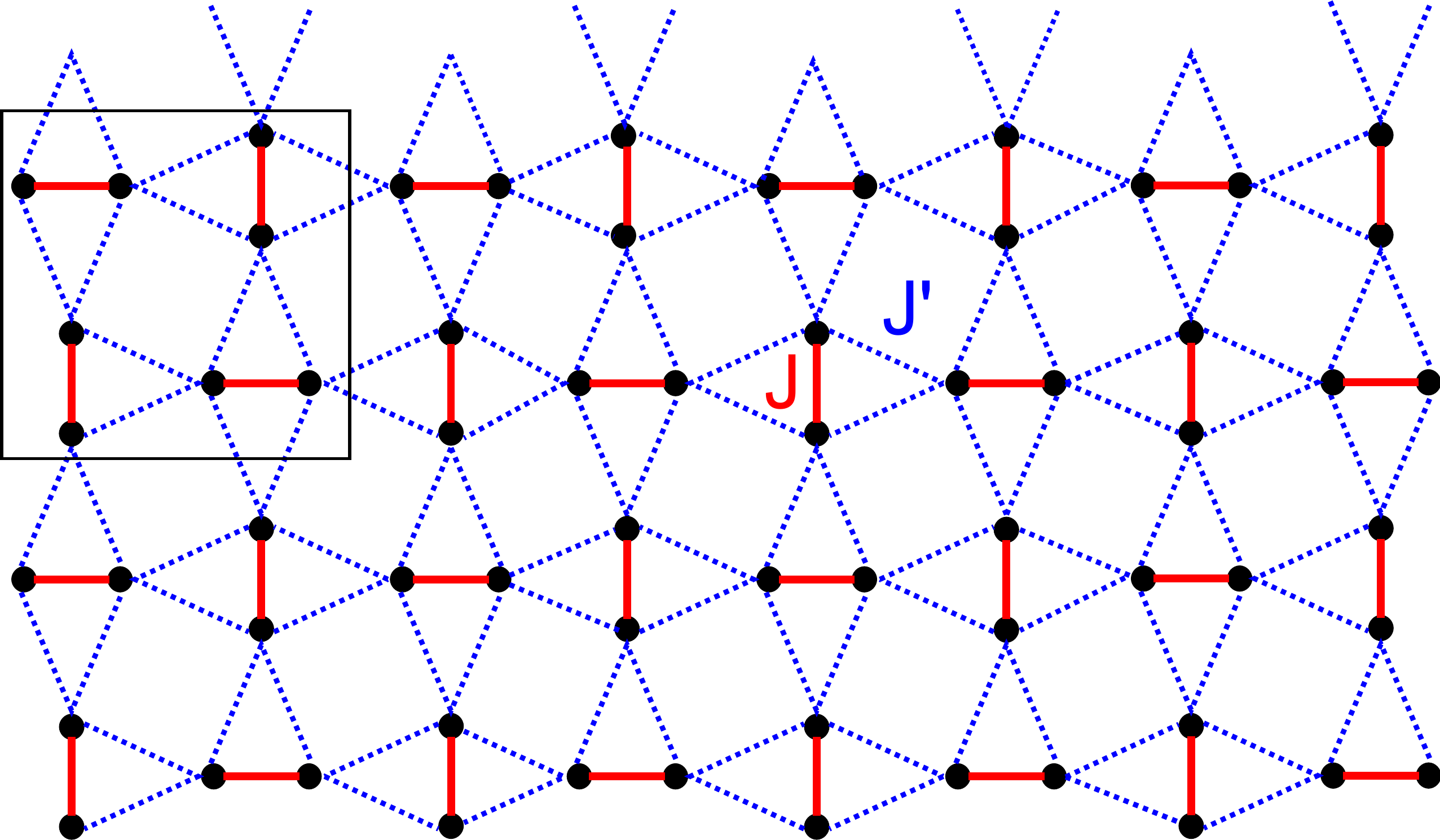}
\caption{Illustration of the four-leg Shastry-Sutherland tube. Solid red bonds denote inter-dimer couplings $J$ and dashed blue lines refers to the intra-dimer coupling $J'$. The tube is periodically coupled in the vertical direction as indicated by the blue dashed lines at the upper end. The unit cell is shown as thin black box covering four dimers.}
\label{pic:model}
\end{figure}

%%%%%%%%%%%%%%
%
%
In Fig.~\ref{pic:model} we show the geometry of the lattice under consideration, which we refer to as a four-leg Shastry-Sutherland tube. On this geometry, we study the spin-$1/2$ Shastry-Sutherland model in an external magnetic field $h$,
%%%%%%%%%%%%%%%%%%%%%%%%%%%%%%%%%%%%%%%%%%%%%%%%%%%%%%%%%%%%%%%%%%%%%%%%%%%%%%%%%%%%%%%%%%%%%%%%%%%%%%%%%%%%%%%%%%%%%%%
\begin{equation}
\mathcal{H} = J \sum\limits_{\ll i,j \gg} \vec{S}_i \cdot \vec{S}_j  + J' \sum\limits_{<i,j>} \vec{S}_{i} \cdot \vec{S}_{j} - h \sum\limits_i S^z_i \quad ,
\label{eq:ham}
\end{equation}
%%%%%%%%%%%%%%%%%%%%%%%%%%%%%%%%%%%%%%%%%%%%%%%%%%%%%%%%%%%%%%%%%%%%%%%%%%%%%%%%%%%%%%%%%%%%%%%%%%%%%%%%%%%%%%%%%%%%%%%
with the bonds $\ll i,j \gg$ building an array of orthogonal dimers and the bonds $<i,j>$ representing inter-dimer couplings. The four-leg Shastry-Sutherland tube has a four-dimer unit cell which is shown in Fig.~\ref{pic:model}.

Here we are not interested in the full phase diagram of Eq.~\eqref{eq:ham}. We restrict ourselves to parameter ratios $J'/J$ for which the four-leg Shasty-Sutherland tube is in the exact product state of singlets for $h=0$, since this is expected to be the relevant coupling regime for SrCu(BO$_3$)$_2$ in two dimensions.\cite{Matsuda13} Indeed, the singlet product state of dimers is, as for the two-dimensional case, an exact eigenstate of the system being the ground state up to $J'/J\approx 0.7$ which can be estimated by extrapolating the one-triplon gap with pCUTs (not shown). The latter value is very close to the one of the two-dimensional Shastry-Sutherland model \cite{Lou12,Corboz13} suggesting that the four-leg tube shares many similarities with its two-dimensional counterpart.  
%
%
%%%%%%%%%%%%%%%%%%%%%%%
%%%%%%%%%%%%%%%%%%%%%%%

% 
%
%%%%%%%%%%%%%%
\section{Methods}
%%%%%%%%%%%%%%
%
In this section, we discuss the pCUTs aiming at the derivation of an effective low-energy model which is treated either by the classical approximation (CA) or by exact diagonalization (ED), and briefly mention details of our DMRG approach.    
%
%%%%%%%%%%%%%%
\subsection{pCUT(+CA/ED)}
%%%%%%%%%%%%%%
%
The pCUT method \cite{Knetter00,Knetter03} has been used successfully for the two-dimensional Shastry-Sutherland model \cite{Dorier08} as well as for the two-leg Shastry-Sutherland tube \cite{Manmana11}. Essentially, the pCUT transforms Eq.~\ref{eq:ham} into an effective model conserving the number of triplons. Triplons with total spin one are the elementary excitations of coupled-dimer systems and can be viewed as triplets dressed with a polarization cloud \cite{Schmidt03}. In a finite magnetic field, the relevant processes have maximum values of total $S_z$. Other channels are only important if bound states of triplons with different quantum numbers become relevant at low energies \cite{Manmana11}. The general form of the effective low-energy model is then given by
\begin{equation}
\mathcal{H}_{\rm eff} = \sum_{i,\delta} t_\delta ^{o}\, b^\dagger_{i+\delta} b^{\phantom{\dagger}}_{i} + \sum_{i,\delta_n} V_{\delta_1,\delta_2,\delta_3}^{o} \, b^\dagger_{i+\delta_3} b^\dagger_{i+\delta_2} b^{\phantom{\dagger}}_{i+\delta_1}b^{\phantom{\dagger}}_{i} \,\ldots\, ,
\label{eq:ham_eff}
\end{equation}
where the sums run over the sites $i$ of the effective square lattice build by dimers of the Shastry-Sutherland model and $o\in\{{\rm v},{\rm h}\}$ gives the orientation {\it vertical} or {\it horizontal} of dimer $i$. The dots "$\ldots$" represent terms containing more than four operators which we do not consider here. The hardcore boson operator $b^\dagger_i$ ($b^{\phantom{\dagger}}_{i}$) corresponds to the creation (annihilation) of a triplet $|t^1\rangle$ on dimer $i$. The amplitudes $t_\delta ^{o}$ and $V_{\delta_1,\delta_2,\delta_3}^{o}$ are obtained as high-order series expansions in $J'/J$ in the thermodynamic limit. We have calculated order 15 for one-body terms $t_\delta$ and order 14 for two-body terms $V_{\delta_1}^{\delta_2,\delta_3}$. In this work we use bare series which are fully converged for $J'/J=0.3$.

In the following we focus on certain aspects of the effective model which are specific to the four-leg Shastry-Sutherland tube and which are different when compared to the two-dimensional case. For more general properties we refer to the literature \cite{Dorier08,Manmana11}

Let us start by discussing the amplitudes $t^{\phantom{\dagger}}_\delta$ of the one-particle operators which are also given in the Appendix \ref{sect:Appendix}. As for the two-dimensional case \cite{Dorier08}, one finds only two types of terms: a chemical potential $\propto b^\dagger_i b^{\phantom{\dagger}}_i$ and a one-particle hopping over the diagonal of the effective dimer square lattice. The chemical potential for the four-leg Shastry-Sutherland tube is slightly different on horizontal and vertical dimers. In fact, the value is lower on vertical dimers. As a consequence, typical magnetization structures found by pCUT have particles dominantly on vertical dimers. But the energy difference in the chemical potentials between horizontal and vertical dimers is very small, e.g.~$10^{-9}J$ for $J'/J=0.3$, since it arises perturbatively only due to a different amplitude in order 10. This energy scale is therefore difficult to resolve by the DMRG. 

Two-particle operators with amplitudes $V_{\delta_1,\delta_2,\delta_3}^{o}$ are dominated by repulsive density-density interactions and correlated hopping terms, i.e.~processes where a particle is only allowed to hop if another particle is present (see also Appendix \ref{sect:Appendix}). Correlated hopping terms arise in order 2 perturbation theory and represent the dominant kinetic processes in the effective Hamiltonian. In contrast to the two-dimensional case, the amplitudes $V_{\delta_1,\delta_2,\delta_3}^{o}$ differ for operators which are related to each other by a 90$^\circ$ rotation due to the finite extension of the four-leg tube in the transverse direction, as was already seen for the one-particle operators. But we stress that these differences are typically tiny, since they originate from virtual fluctuations which wind around the tube and therefore only show up in high orders of the perturbation. As a consequence, we use in figures and when appropriate the notation $\mu\equiv t_{(0,0)}^{\rm v}\approx t_{(0,0)}^{\rm h}$ for the chemical potential on all dimers and, in the same spirit, $V_n$ for the repulsive density-density interactions which have been introduced in Ref.~\onlinecite{Dorier08} for the 2D Shastry-Sutherland model (see also Appendix \ref{sect:Appendix}).

Altogether, the effective model of the four-leg Shastry-Sutherland tube is very similar to the one for the two-dimensional case \cite{Dorier08}. However, this does not imply that the solution of the effective model is the same in both cases. Especially at low magnetizations, unit cells of magnetization plateaus become large and the finite extension of the tube can matter.  

The derivation of the effective model by pCUTs is only the first step, since the solution of Eq.~\eqref{eq:ham_eff} is by far not simple. Here we apply two strategies to tackle the effective model: i) a CA along the lines of Ref.~\onlinecite{Dorier08} and ii) ED. In the following we call these two approaches pCUT(+CA) and pCUT(+ED).

i) For the CA one applies a Matsubara-Matsuda transformation \cite{Matsubara56} to rewrite Eq.~\eqref{eq:ham_eff} in terms of effective pseudo-spins 1/2. Then, spin 1/2 operators are replaced by classical vectors yielding a classical energy which is minimized for a set of unit cells. We considered unit cells up to $8\times 4$ dimers. Physically, this approach works fine as long as quantum fluctuations are not too strong, i.e.~it is exact in the limit where diagonal density-density interactions dominate over kinetic processes. This is indeed the case if one only considers one-particle hopping processes. In contrast, large correlated hopping processes are not treated well by the CA which leads to a breakdown of this approach as shown below.
 
ii) We have used ED to diagonalize the effective model (\ref{eq:ham_eff}) on finite clusters. This is done with the Lanczos algorithm \cite{Lanczos50} allowing us to treat systems with up to 32 dimers and with up to $M=1/3$ using either open or periodic boundary conditions. Let us stress that the ED of the effective model has less finite-size effects compared to an ED (or DMRG) of the original model on the same cluster, since the effective model has been derived in the thermodynamic limit. 

%
%%%%%%%%%%%%%%
\subsection{pCUT$_{\rm finite}$(+ED)}

Alternatively, one can also i) derive the effective low-energy model by pCUTs directly on {\it finite} clusters and then ii) solve this cluster-dependent effective model by ED. In the following we denote this approach by pCUT$_{\rm finite}$(+ED). This allows a straightforward comparison with DMRG results obtained on the same cluster. 

In contrast to Eq.~\eqref{eq:ham_eff}, the amplitudes of the effective model do depend directly on the absolute location of particles on the cluster under study
\begin{equation}
\mathcal{H}_{\rm eff}^{L} = \sum_{i,\delta} t_{i,\delta} ^{o}\, b^\dagger_{i+\delta} b^{\phantom{\dagger}}_{i} + \sum_{i,\delta_n} V_{i,\delta_1,\delta_2,\delta_3}^{o} \, b^\dagger_{i+\delta_3} b^\dagger_{i+\delta_2} b^{\phantom{\dagger}}_{i+\delta_1}b^{\phantom{\dagger}}_{i} \,\ldots\, ,
\label{eq:ham_eff_finite}
\end{equation}
where $\mathcal{H}_{\rm eff}^{L}$ refers to the effective model on a cluster of size $L\times 4$. The dependence of the amplitudes $t_{i,\delta} ^{o}$, $V_{i,\delta_1,\delta_2,\delta_3}^{o}$, etc.~on the variable $i$ corresponds to the present finite-size effects. In this work we have calculated all amplitudes of the effective model including all multi-particle processes up to order 10 in $J'/J$ and up to $L=8$ ($L=6$) for $M=1/8$ ($M=1/4$). As for pCUT(+ED), we use bare series which are fully converged for $J'/J=0.3$.

There can be important differences between pCUT$_{\rm finite}$(+ED) and pCUT(+ED) which we would like to discuss shortly for the chemical potential $\propto b^\dagger_{i} b^{\phantom{\dagger}}_{i}$. In contrast to pCUT(+ED) discussed above, the chemical potential varies for different dimers on the cluster. In order two, the chemical potential $J-(J'/J)^2$ is still the same on all horizontal and vertical dimers except for the vertical dimers located at the edge of the cluster. Here one finds $J-1/2(J'/J)^2$. The different order-two amplitude originates from the fact that a triplet located on a vertical dimer on the left (right) edge of the cluster cannot perform a virtual fluctuation to the left (right) in contrast to dimers in the inside as well as horizontal dimers on the edge which still can fluctuate up and down. This difference in the chemical potential is large compared to the one between vertical and horizontal dimers in the effective model in the thermodynamic limit arising from the tube geometry. As a consequence, the magnetization profiles as well as the ground-state energy obtained by pCUT(+ED) and the ones obtained by pCUT$_{\rm finite}$(+ED) and DMRG might differ.  

%%%%%%%%%%%%%%
%
%
\begin{figure}
\includegraphics[width=0.79\columnwidth]{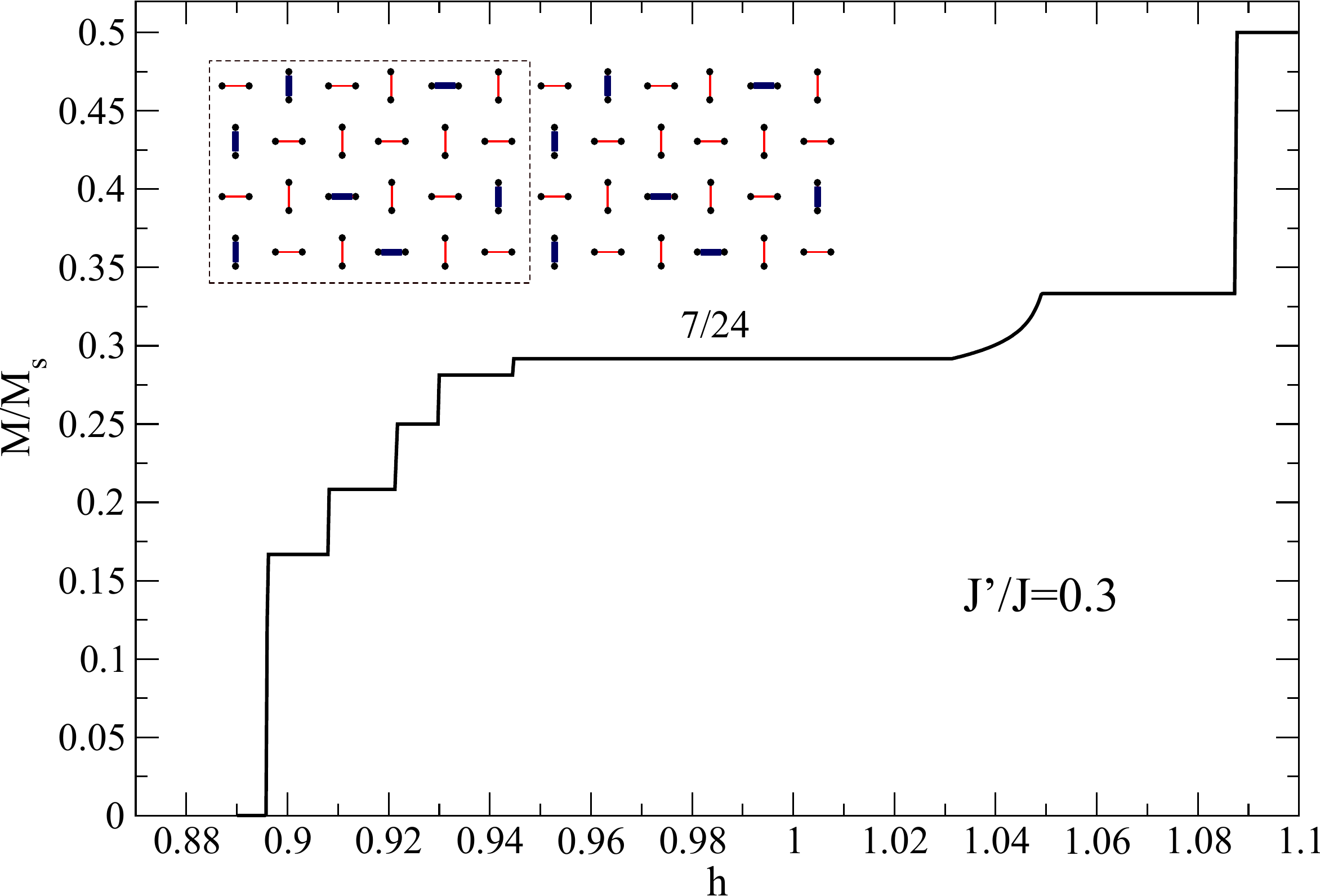}\vspace*{+2mm}
\includegraphics[width=0.79\columnwidth]{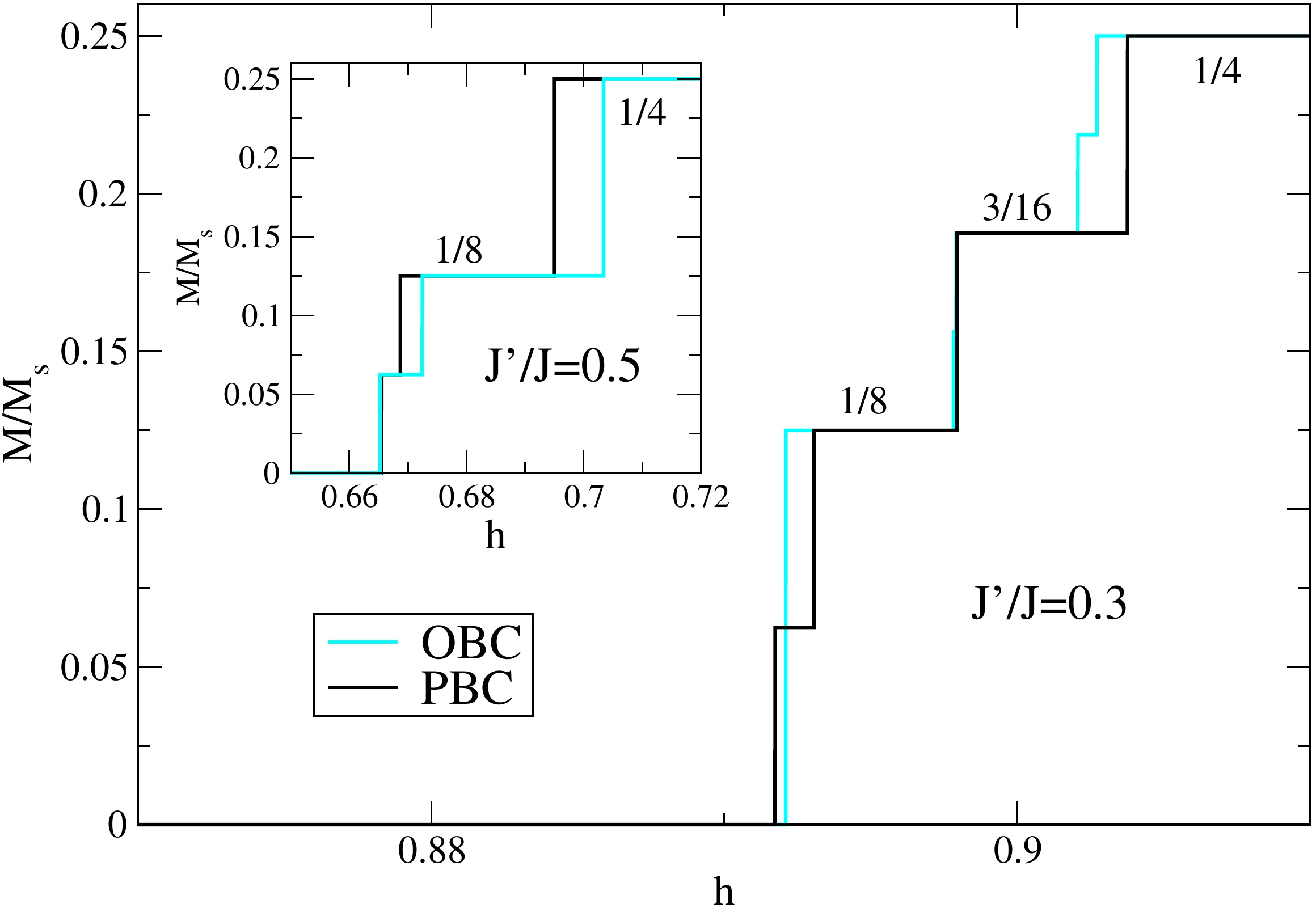}
\includegraphics[width=0.9\columnwidth]{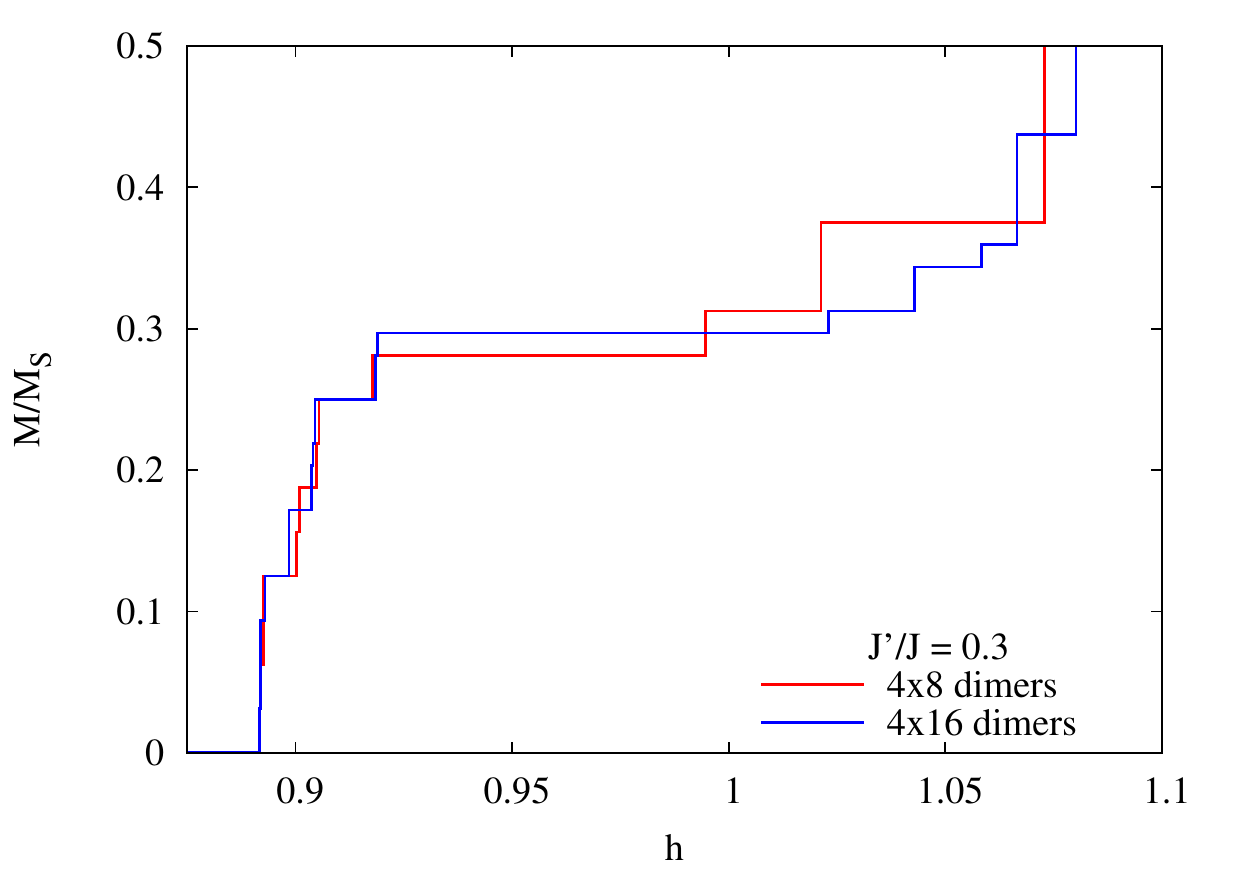}
\includegraphics[width=0.9\columnwidth]{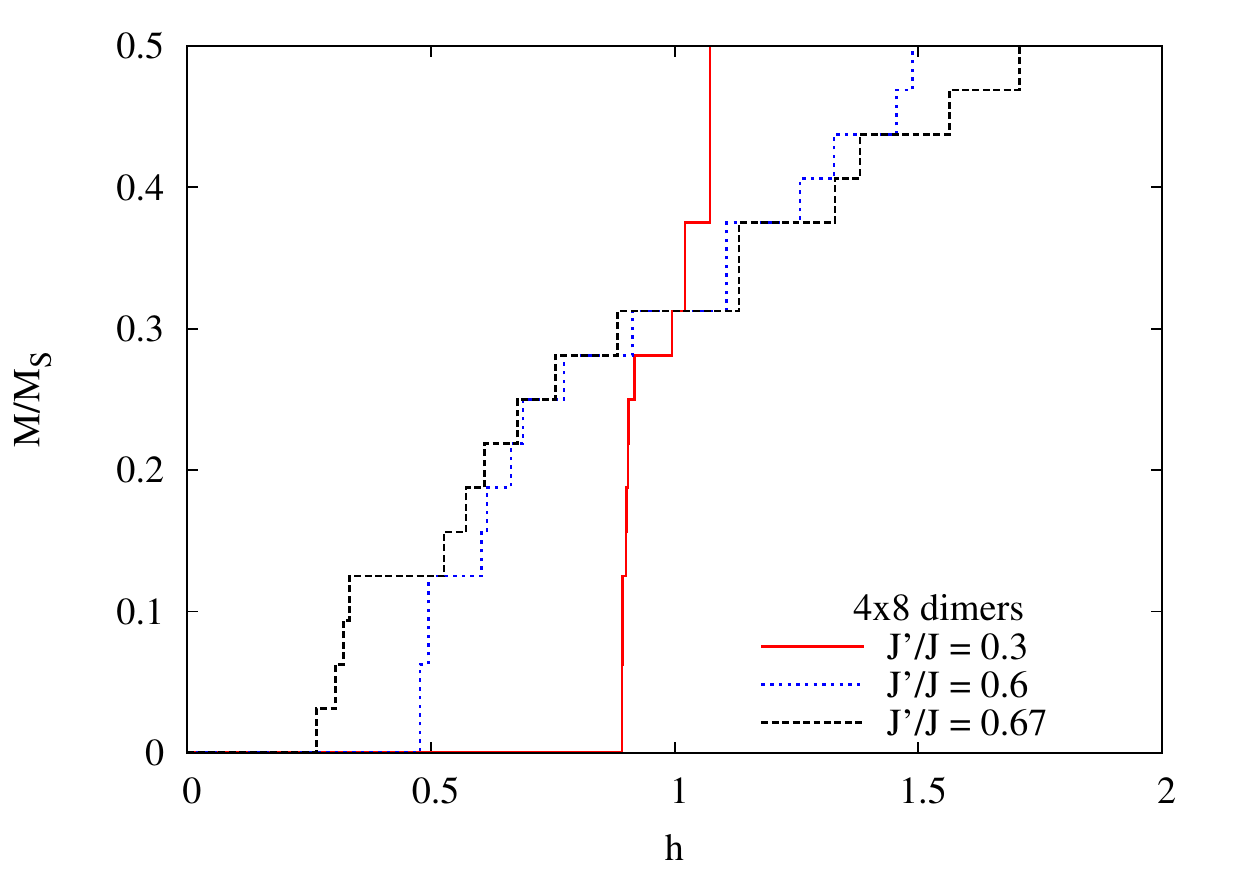}
\caption{Magnetization $M/M_{\rm s}$ as a function of the magnetic field $h$  obtained with pCUT(+ED) and DMRG. {\it First panel}: Results from pCUT(+CA). Inset shows magnetization profile at $M=7/24$. {\it Second panel}: Results from pCUT(+ED) on a $8\times 4$ with periodic (PBC) and open (OBC) boundary conditions for $J'/J=0.3$ as well as $J'/J=0.5$ in the inset. {\it Third panel}: DMRG results for systems with $8\times 4$ and $16\times 4$ dimers for $J'/J = 0.3$. {\it Fourth panel}: DMRG results for systems with $8\times 4$ dimers for various values of $J'/J$.}
\label{fig:mags}
\end{figure}
%
%%%%%%%%%%%%%%
\subsection{DMRG}

In most cases we treat systems with up to 64 dimers using open boundary conditions in the longitudinal direction, and in particular cases we go up to 144 dimers. 
Typically, we use 35 sweeps keeping up to $m=1500$ density matrix eigenstates, leading to a discarded weight of the order of $10^{-8}$ or better; in some cases which appear more difficult to converge, we apply up to 50 sweeps and $m=2500$. 
The results on different system sizes are consistent with each other, so that we refrain from performing a more detailed study varying the sweeping procedure etc. as discussed in Refs.~\onlinecite{WhitePRL2007,WhiteScienceKagome,Stoudenmire2D}. 
We also do not perform a finite-size scaling procedure in terms of the cylinder circumference as performed, e.g., in Ref.~\onlinecite{WhiteScienceKagome} for kagome systems. 
Strictly speaking, we therefore cannot make predictions for the behaviour in the true 2D case. 
Instead, our strategy here is to focus on systems with four legs and by carefully comparing energies as well as local observables to pCUTs identify the mechanisms leading to the interesting plateaus discussed in this paper. 
Note that in our previous study Ref.~\onlinecite{Matsuda13}, DMRG results for 2D behaviour when applying PBC where discussed. 
There, we focused on ground-state energies at finite magnetic fields, which are much easier to obtain and so could lead to insights into the 2D behavior. 
Here, we go beyond and obtain accurate results for local observables, however, restricting us to the aforementioned configurations.  
% 
%
%%%%%%%%%%%%%%
\section{Magnetization curves}
%%%%%%%%%%%%%%
%
%
In Fig.~\ref{fig:mags} we present our results for the magnetisation curves obtained with DMRG (pCUT(+ED)) for finite systems with up to 64 (32) dimers for various values of $J'/J$. One observes plateaus at $M=1/8$, $M=1/4$, and $M=1/2$ with both methods and for different system sizes. Most remarkably, the plateau at $M=1/8$ is very stable and grows with increasing $J'/J$. 

In contrast to the 2D case, there is no prominent plateau with $M=1/3$. This is well understood in terms of the effective pCUT model, since the corresponding 2D structure consisting of diagonal stripes on alternatively vertical and horizontal dimers is energetically not favored on the four-leg tube, since one has to pay large repulsive interactions. 

Additionally, there are many other plateau structures which we mostly attribute to finite-size artifacts, since the associated magnetization profile is not corresponding to any regular structure and therefore specific for a given cluster. One exception might be the existence of a plateau around $M\sim 0.3$. Here DMRG displays a broad magnetization plateau. Additionally, we observe a very stable plateau at $M=7/24$ within pCUT(+CA) containing similar features in the magnetization profiles (see Fig.~\ref{fig:mags}). Such a plateau has likely a conventional structure corresponding to a classical solution of the effective model and we therefore do not focus on this plateau in the following.

We will now focus on the sequence $M=1/8$, $M=1/4$, and $M=1/2$ where all our techniques display very regular structures. The magnetization profile of the plateau at $M=1/2$ is just the known and very well understood one appearing also in the 2D Shastry-Sutherland model. Here the magnetization is dominantly on one sublattice, which is a consequence of the very large repulsive interaction $V_1$ between nearest-neighbor dimers. Consequently, we do not focus on this plateau but we concentrate on the specific magnetizations \mbox{$M=1/8$} and $M=1/4$ for which pCUT and DMRG allow a consistent and interesting interpretation. Afterwards, we discuss the intermediate regime between $M=1/8$ and $M=1/4$.
% 
%
%%%%%%%%%%%%%%
\subsection{1/8 plateau}
%%%%%%%%%%%%%%
%
%
A very robust structure found in DMRG is the one at $M=1/8$. In the following we analyze first this magnetization for a rather small value $J'/J=0.3$ where the effective model derived by pCUT is fully converged. Here we compare our findings by DMRG, pCUT(+CA/+ED), and pCUT$_{\rm finite}$(+ED) and we give the physical origin of the observed structure. Afterwards, we discuss the properties of the 1/8 plateau as a function of $J'/J$. 

%
%%%%%%%%%%%%%%
\begin{figure}
\includegraphics[width=\columnwidth]{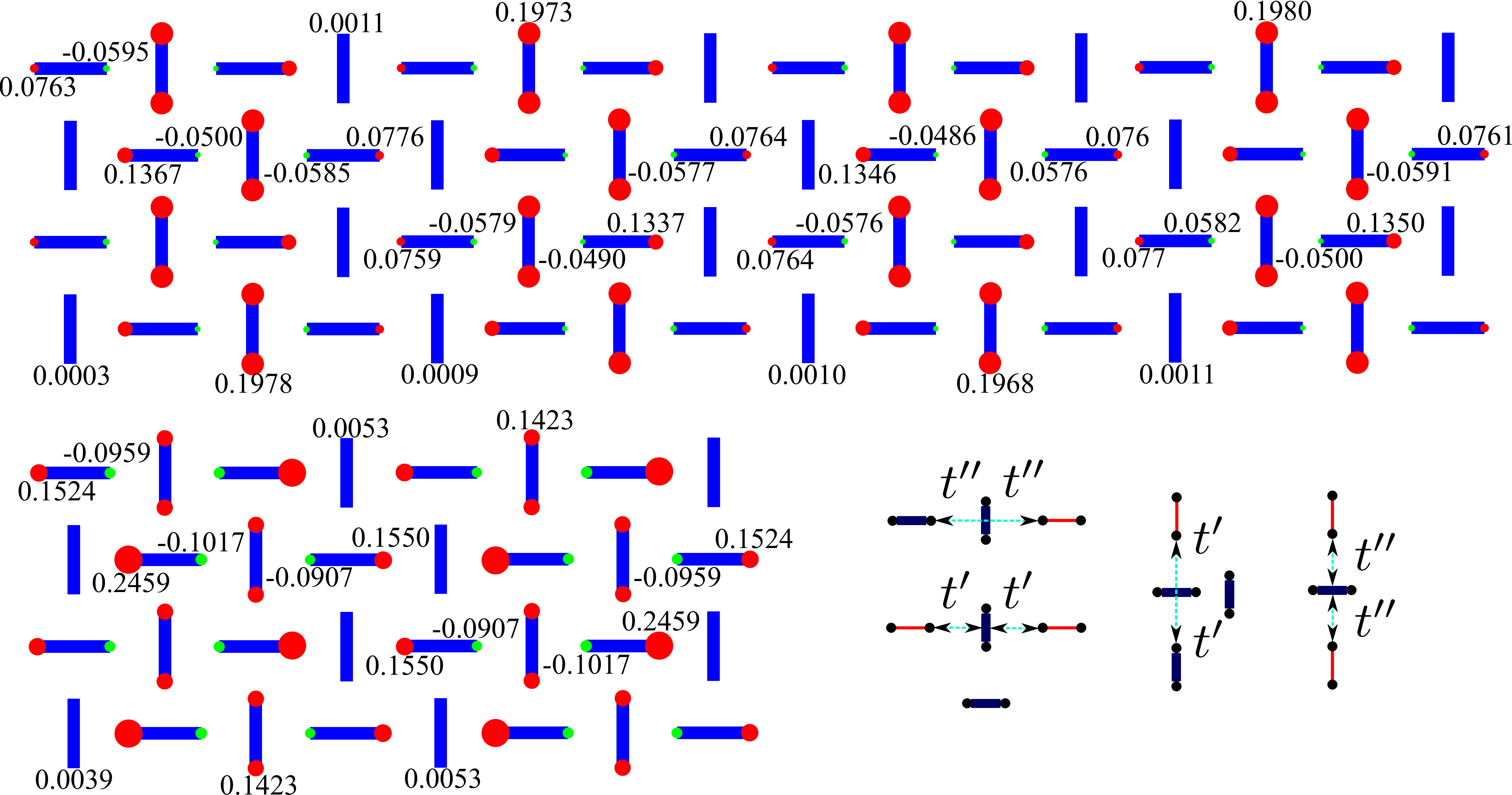}
\caption{Local magnetization of the 1/8 plateau using DMRG. Upper (lower left) panel corresponds to an open tube with 64 dimers (32 dimers) and $J'/J=0.3$ ($J'/J=0.6$). The diameters of circles relate to the local magnetization which is also given explicitly as numbers. Positive (negative) values denote a local magnetization parallel (antiparallel) to the field direction. Lower right panel gives an illustration of the dominant order-two correlated hopping processes $t'$ and $t''$.}
\label{fig:0.125_DMRG}
\end{figure}
%%%%%%%%%%%%%%
% 

For the open tube with 64 dimers as shown in Fig.~\ref{fig:0.125_DMRG} for $J'/J=0.3$, DMRG finds a regular pattern of four structures which we name {\it wheels}. In a wheel the magnetization is uniformly distributed among the four vertical dimers around the tube so that each dimer contains roughly half of the magnetization of a single triplet. The total value of $S^z$ in a wheel is therefore 2. Between wheels there are two rows of dimers where almost no magnetization is present leading to a unit cell of 16 dimers. The same kind of structure is also found for open tubes with 16, 32, and 48 dimers having 1,2, and 3 wheels. It is therefore very certain that this magnetization plateau at $M=1/8$ is present in the thermodynamic limit. The energy per site is fully converged when comparing different clusters and one obtains \mbox{$\epsilon_0^{\rm 1/8}=-0.31924\,J$} for $J'/J=0.3$. 

\begin{figure}
\includegraphics[width=0.97\columnwidth]{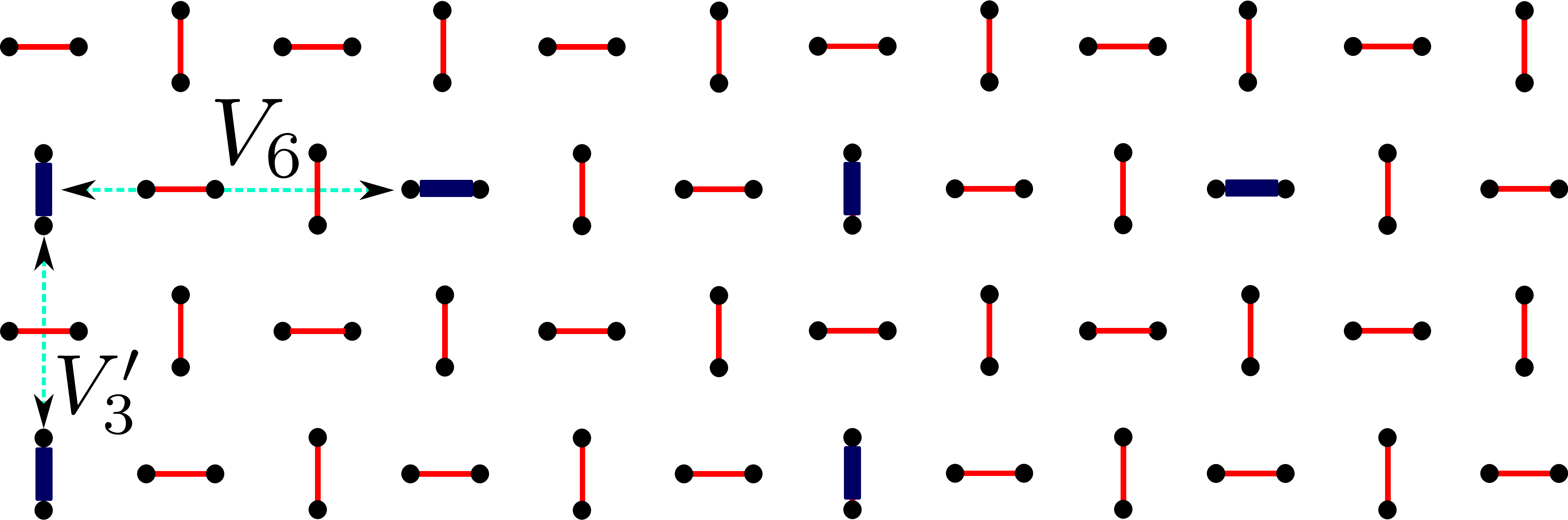}
\caption{Illustration of the plateau with $M=1/8$ obtained by pCUT(+CA). Thin red (thick dark) lines correspond to empty (filled) dimers. The most relevant two-particle interactions $V_3^\prime$ and $V_6$ are shown as grey/cyan dashed lines with arrows on both sides pointing to the two involved filled dimers.}
\label{fig:0.125_CA}
\end{figure}

One should stress that the detected structure of the local magnetization is very surprising. Naively, one expects magnetization plateaus at low magnetizations which correspond to crystals of single-triplon excitations being stabilized by repulsive density-density interactions. One such example can be seen with pCUT(+CA) displayed in Fig.~\ref{fig:0.125_CA} for $J'/J=0.3$.

The classical structure with a unit cell of 24 dimers has the energy per site $\epsilon_{\rm 0, cl}^{\rm 1/8}=-0.31901\,J$ at $J'/J=0.3$, i.e.~the energy is $\sim 2\cdot 10^{-4}\,J$ higher compared to the one found by DMRG. The classical structure results from minimizing density-density interactions, since it avoids paying all repulsive interactions below order 6 as for the classical structures for the 2D case \cite{Dorier08}. In fact, it is interesting to realize that the classical 1/8 structure found for the four-leg tube is similar in spirit to the realized 1/9 plateau in 2D \cite{Dorier08}.

Next we add quantum fluctuations on top of the classical solution to show that the energy difference to the DMRG result cannot be explained this way. Quantum fluctuations are induced in the effective model by kinetic processes $\tau_j$ like hopping, correlated hopping, etc. . The leading correction to the classical energy per site is calculated by summing over all contributions $-\tau_j^2/\Delta E_j$ where $\Delta E_j$ is the energy difference between the energy of the intermediate state after acting with the kinetic process $\tau_j$ on the classical ground state and the energy $E_{\rm 0, cl}$ of the classical state itself. This approach is fully controlled if $\tau_j^2\ll\Delta E_j$ for all $j$. For the rather small value $J'/J=0.3$ considered here, this is exactly the case, since the leading perturbative order in $J'/J$ involved in all $\Delta E_j$ is lower than the coresponding one in $\tau_j^2$. 

If one adds quantum fluctuations on top of the classical structure, one finds that kinetic processes are never able to explain the observed energy difference between DMRG and the CA of the effective model, since the distance between particles in the classical structures is already so large that kinetic processes are not able to reduce the energy sufficiently. Additionally, also the structure made of single dressed triplons is clearly incompatible with the magnetization profile deduced by DMRG. In conclusion, the pCUT(+CA) is not able to reproduce the findings of DMRG even for values of $J'/J$ where the effective model is fully converged.

%
%%%%%%%%%%%%%%
%%%%%%%%%%%%%%
% 

\begin{figure}
\includegraphics[width=\columnwidth]{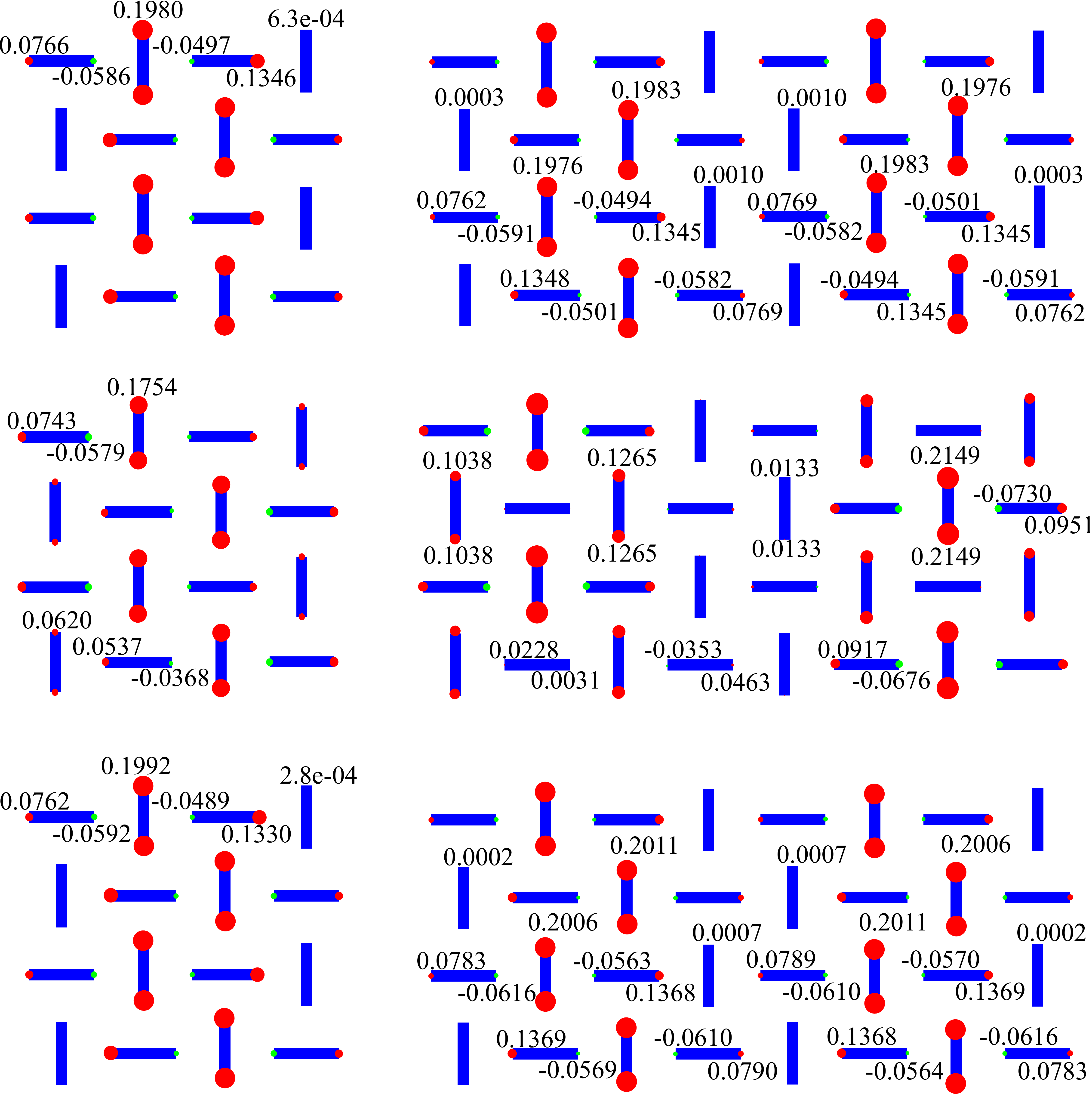}
\caption{Local magnetization of the 1/8 plateau on an open tube with 16 (left) 32 (right) dimers for $J'/J=0.3$ using DMRG (upper panel), pCUT(+ED) (middle panel), and pCUT$_{\rm finite}$(+ED) (lower panel). The diameters of circles relate to the local magnetization which is also given explicitly as numbers. Positive (negative) values denote a local magnetization parallel (antiparallel) to the field direction. One observes the realization of one (left) and two (right) quantum wheels.}
\label{fig:0.125_pCUT_ED}
\end{figure}

Therefore, it cannot be the (fully converged) effective model derived by pCUTs which is problematic, but it has to be the CA which fails. The CA is not able to treat the correlated hopping processes well as we show below. This becomes clear when solving the effective model by ED on finite clusters so that quantum fluctuations are taken into account exactly. The corresponding results for $J'/J=0.3$ on clusters with 16 and 32 dimers are shown in Fig~\ref{fig:0.125_pCUT_ED}. Most notably, the pCUT(+ED) and pCUT$_{\rm finite}$(+ED) are in agreement with DMRG: one finds a regular pattern of wheels and the ground-state energy per site $-0.31924\,J$ on an open cluster with 16 dimers is identical. In fact, the local magnetization profile of pCUT$_{\rm finite}$(+ED) and DMRG, which are performed on the same cluster, are quantitatively the same. For pCUT(+ED), where the ED is done on the effective model evaluated in the thermodynamic limit, one observes the same pattern but more magnetization is found at the edge of the cluster. This is mainly due to the isotropic chemical potential which is the same on all dimers in contrast to the finite-size approaches as explained in the method section.

What is the physical origin of these wheels? Since the wheels are (i) already stable for small values of $J'/J$ and (ii) not realized in the CA, this structure has to come from the perturbatively leading kinetic processes in the effective pCUT model. The latter are correlated hopping terms arising in order two perturbation theory, i.e.~wheels are two-particle objects. 

Let us consider one pair of particles. In order to profit from the leading correlated hopping processes $t'$ and $t''$ (see also Appendix \ref{sect:Appendix} and Fig.~\ref{fig:0.125_DMRG}), particles should be either next-nearest neighbors in a diagonal configuration ($t'$) or nearest neighbors ($t'$ and $t''$). The first configuration is by far more attractive, since one does not need to pay the large nearest-neighbor repulsion $V_1$ arising in leading order perturbation theory. In contrast, the repulsive density-density interaction over the diagonal $V_2$ is a third-order process and therefore small as long as $J'/J$ is small. Nevertheless, this interaction is larger than the ones appearing in the classical structures and a classical plateau of particles being in a diagonal configuration does not represent the classical energy minimum.      

%
%%%%%%%%%%%%%%
%%%%%%%%%%%%%%
% 
One therefore has to maximize quantum fluctuations due to correlated hopping. This is done by taking the four low-energy states with energy $E^{\rm l}_{1/8}=2\mu+V_2$ where two particles are next-nearest neighbors on adjacent rows. Each of the four states can fluctuate to two states where both particles are nearest neighbors with energy $E^{\rm h}_{1/8}=2\mu+V_1$ due to the correlated hopping $t'$. In total, this leads to the finite matrix 
\begin{align}
 \begin{pmatrix}
 E^{\rm l}_{1/8} & 0 & 0 & 0 & + t' & + t' & 0 & 0 \\
 0 & E^{\rm l}_{1/8}  & 0 & 0 & 0 & 0 & + t' & + t' \\
 0 & 0 & E^{\rm l}_{1/8}  & 0 & 0 & - t' & - t' & 0 \\
 0 & 0 & 0 & E^{\rm l}_{1/8}  & - t' & 0 & 0 & - t' \\
 + t' & 0 & 0 & - t' & E^{\rm h}_{1/8}  & 0 & 0 & 0 \\
 + t' & 0 & - t' & 0 & 0 & E^{\rm h}_{1/8}   & 0 & 0 \\
 0 & + t' & - t' & 0 & 0 & 0 & E^{\rm h}_{1/8}   & 0 \\
 0 & + t' & 0 & - t' & 0 & 0 & 0 & E^{\rm h}_{1/8}   \\
 \end{pmatrix}% \\ &= 
\end{align}
and the resulting ground-state energy of a single wheel is obtained analytically as
\begin{eqnarray}
 E_{\rm wheel} &=& \frac{E^{\rm l}_{1/8} + E^{\rm h}_{1/8}}{2} - \frac{1}{2} \sqrt{ \left( E^{\rm h}_{1/8} - E^{\rm l}_{1/8} \right)^2 + 16 {t'}^{2} }\nonumber\\
      &=& 2\mu+\frac{V_1 + V_2}{2} - \frac{1}{2} \sqrt{ \left( V_1-V_2 \right)^2 + 16 {t'}^{2} } \quad .
\end{eqnarray}
The corresponding eigenvector is a highly entangled two-particle bound state. Both particles are with the same probability in one of the four low-energy states and the density is approximately 1/2 on all four vertical dimers inside the wheel for $J'/J=0.3$ just as observed in the magnetization profile deduced by DMRG. 

If one assumes the interactions between wheels to be zero (which is almost the case as shown below), then the ground-state energy per spin of the crystals of wheels is perturbatively
\begin{eqnarray}
 \frac{\epsilon_{0,{\rm wheel}}^{\rm 1/8}}{J} &=& E_{\rm wheel}/32-\frac{3}{8}\nonumber\\
      &\approx & -\frac{5}{16}-\frac{1}{16}\left(\frac{J'}{J}\right)^2-\frac{5}{128}\left(\frac{J'}{J}\right)^3 \quad .
\end{eqnarray}
Interestingly, one finds the energy per spin \mbox{$\epsilon_{0,{\rm wheel}}^{\rm 1/8}\approx -0.31918\, J$} for $J'/J=0.3$ in good agreement with DMRG. Note that we have checked that all other quantum fluctuations do not alter this result significantly. Most importantly, the order-three expansion of $\epsilon_{0,{\rm wheel}}^{\rm 1/8}$ remains correct in the thermodynamic limit for the full crystal of wheels.

The astonishing agreement between the extrapolated energy from DMRG and the single-wheel calculation presented above is only possible if the interaction between wheels of the 1/8 plateau is negligible. This is indeed the case: First, the ground-state energy per spin in DMRG of $\nu$ wheels is almost identical to $\nu$ times the single-wheel energy of a 16-dimer cluster which demands the interactions between wheels to be close to zero. Second, assuming the density of particles on verticle dimers in a wheel to be roughly 1/2, the interactions between two neighboring wheels to be paid only involve repulsive density-density interactions in the effective model starting at least in order 6 perturbation theory giving a neglible energy cost.  

Altogether, we found quantitative agreement between all approaches used. The 1/8 plateau consists of almost decoupled wheels in a 16-dimer unit cell. Each wheel is a two-particle bound state which is stabilized by correlated hopping. It is therefore reasonable that the crystal of wheels becomes more stable with increasing $J'/J$, since the delocalization of the two particles due to correlated hopping is even enhanced along the wheels. This is confirmed by DMRG (see also Fig.~\ref{fig:0.125_DMRG}). One observes well defined wheels in a broad range of parameters $J'/J$ and the width of the 1/8 magnetization plateau increases for large values of $J'/J$.    
% 
%
%%%%%%%%%%%%%%t
\subsection{1/4 plateau}
\begin{figure}
\includegraphics[width=0.85\columnwidth]{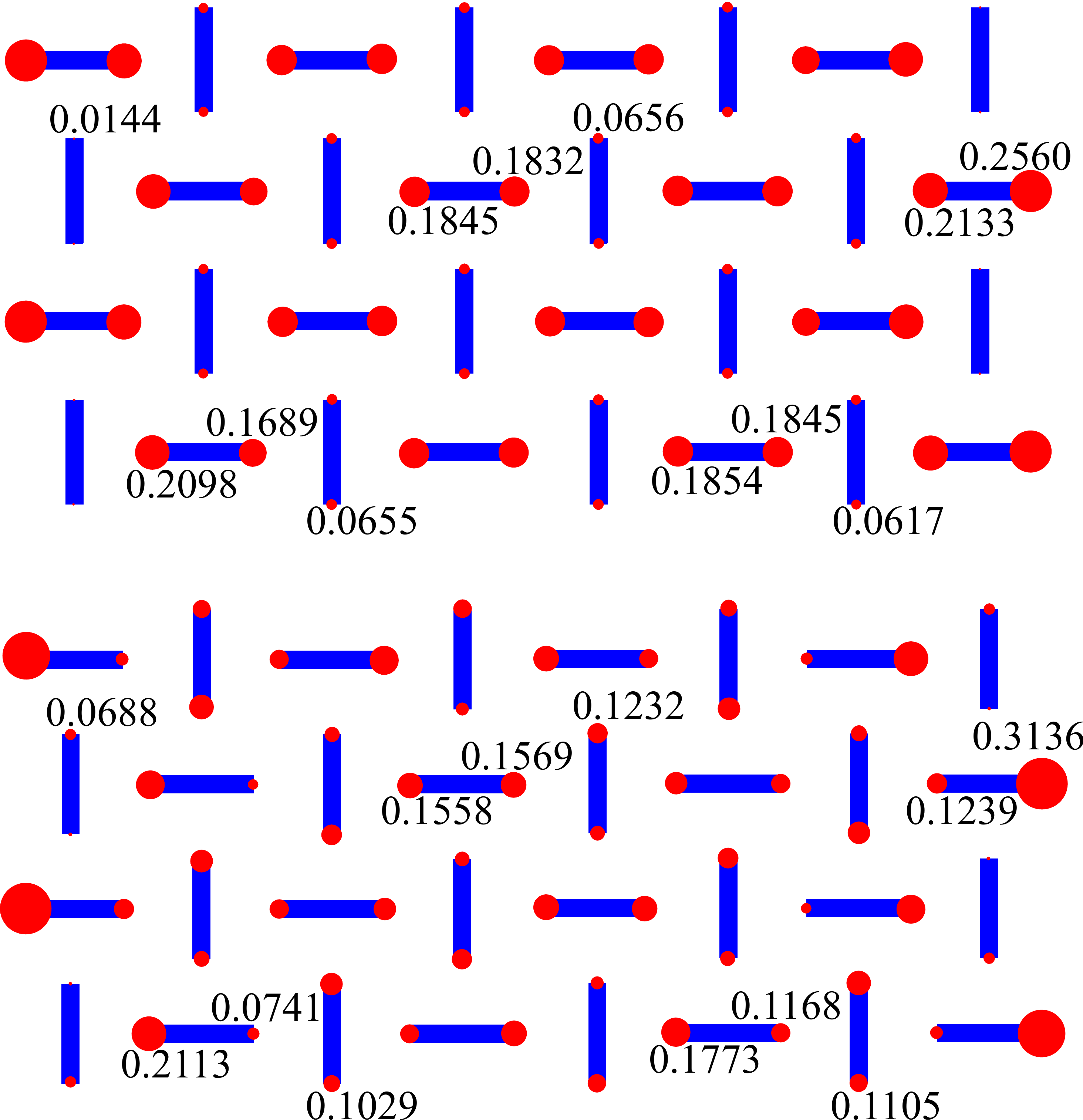}
\caption{Local magnetization of the 1/4 plateau on an open tube with 64 dimers using DMRG. One observes the superposition of four diagonal stripes. Upper (lower) panel corresponds to $J'/J=0.3$ ($J'/J=0.6$). The diameters of circles relate to the local magnetization which is also given explicitly as numbers. Positive (negative) values denote a local magnetization parallel (antiparallel) to the field direction. }
\label{fig:0.25_DMRG}
\end{figure}

%%%%%%%%%%%%%%
%
Next we discuss the second prominent plateau of the four-leg Shastry-Sutherland model which is the one at $M=1/4$. The DMRG finds consistently a magnetization profile shown in Fig.~\ref{fig:0.25_DMRG} for an open cluster with 64 dimers for $J'/J=0.3$, where the magnetization is almost distributed uniformly among all horizontal dimers. The corresponding energy per site is \mbox{$\epsilon_0^{\rm 1/4}=-0.26281\,J$} for $J'/J=0.3$. If one increases the ratio $J'/J$ to 0.6, the overall structure is similar except that the magnetization is transferred more prominently from horizontal to vertical dimers. 
%
%
%%%%%%%%%%%%%%
%%%%%%%%%%%%%%
% 

\begin{figure}
\includegraphics[width=0.97\columnwidth]{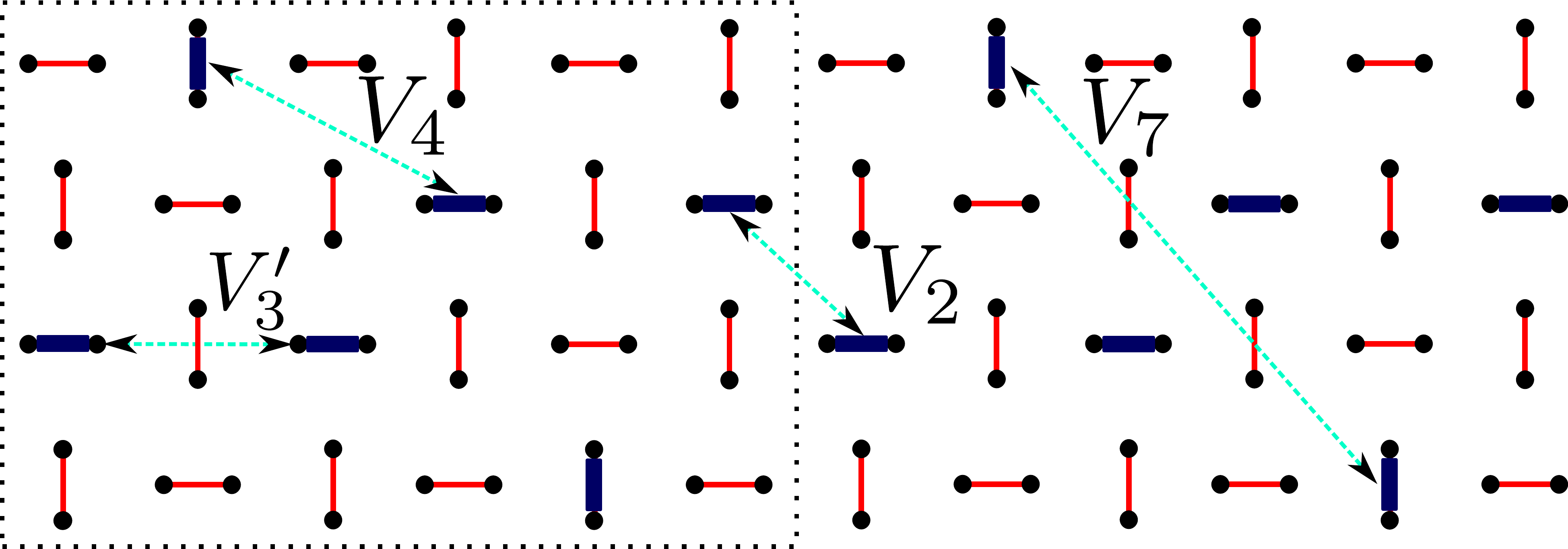}
\caption{Illustration of the plateau with $M=1/4$ obtained by pCUT(+CA). Thin red (thick dark) lines correspond to empty (filled) dimers. The most relevant two-particle interactions $V_2$, $V_4$, $V_3^\prime$, and $V_7$ are shown as grey/cyan dashed lines with arrows on both sides pointing to the two involved filled dimers.}
\label{fig:0.25_CA}
\end{figure}

As for the 1/8 plateau, the DMRG result is not explained within pCUT(+CA) which is shown in Fig.~\ref{fig:0.25_CA}. One finds a classical structure with a 24-dimer unit cell where single particles are far apart in order to minimize repulsive interactions. Its energy per spin is \mbox{$\epsilon_{\rm 0, cl}^{\rm 1/4}=-0.26222\,J$ at $J'/J=0.3$} and therefore \mbox{$\sim 6\cdot 10^{-4}\,J$} higher in energy compared to DMRG. In contrast to the classical 1/8 structure, the one at 1/4 involves two density-density interactions of type $V_2$ per unit cell and quantum fluctuations induced by the dominant order-two correlated hopping $t'$ can overcome half of this energy difference. Nevertheless, there remain discrepancies with respect to energy and with respect to the magnetization profile which cannot be resolved. Consequently, there must a different structure behind the observed 1/4 plateau.
 
%
%%%%%%%%%%%%%%
%%%%%%%%%%%%%%
% 

\begin{figure}
\includegraphics[width=0.8\columnwidth]{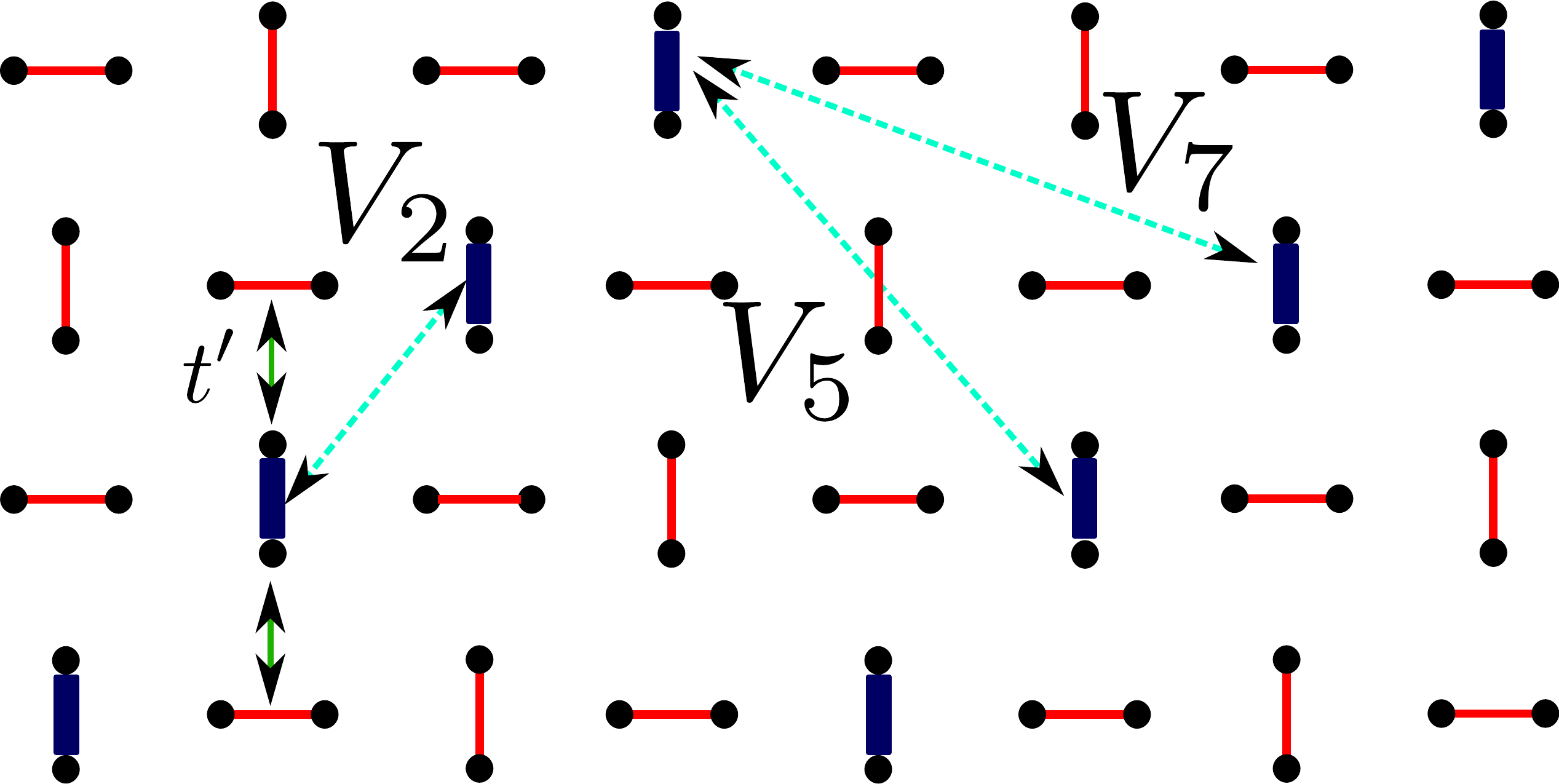}
\caption{Illustration of one semi-classical stripe on vertical dimers relevant for $M=1/4$. Thin red (thick dark) lines correspond to empty (filled) dimers. The most relevant two-particle interactions $V_2$ and $V_5$ are shown as grey/cyan dashed lines with arrows on both sides pointing to the two involved filled dimers. Each particle of the stripe can fluctuate by the dominant order-two correlated hopping process $t'$ to the closest horizontal dimers above and below the stripe which is sketched for the particle in the second row.}
\label{fig:0.25_SemiCA}
\end{figure}

In the following we show that the 1/4 plateau corresponds to a diagonal stripe which winds around the tube as illustrated in Fig.~\ref{fig:0.25_SemiCA}, i.e.~the translational symmetry is broken in the thermodynamic limit and a single stripe is realized which is dressed by quantum fluctuations. We call such a structure semi-classical, since its energy represents a local minimum within pCUT(+CA) which becomes the true ground state when adding leading quantum fluctuations inside the effective pCUT model. The particles are placed on vertical dimers in the thermodynamic limit, since the chemical potential is slightly lower compared to horizontal dimers. Such a diagonal stripe costs dominantly one interaction $V_2$ for each particle and its classical energy per spin $-0.26175\, J$ is higher than $\epsilon_{\rm 0, cl}^{\rm 1/4}$. 

Interestingly, quantum fluctuations induced by correlated hopping lower the classical energy of the diagonal stripe significantly such that this structure is indeed realized in the four-leg Shastry-Sutherland model. Each particle on a vertical dimer of a stripe can use the dominant order-two correlated hopping process $t'$ to fluctuate to the horizontal dimers below and above the stripe. Again, in the intermediate state one has to pay the interaction $V_1$ for particles being nearest neighbors. Summing over these fluctuation channels in order ${t'}^2/\Delta E$, as for the 1/8 plateau, one gets an energy reduction per spin of approximately \mbox{$-8{t'}^2/32 V_1=-(J'/J)^3/32$}, since there are 8 fluctuation channels in the 32-spin unit cell. This results in a semi-classical energy per spin 
\begin{eqnarray}
 \frac{\epsilon_{0,{\rm stripe}}^{\rm 1/4}}{J} &\approx & -\frac{1}{4}-\frac{1}{8}\left(\frac{J'}{J}\right)^2-\frac{15}{256}\left(\frac{J'}{J}\right)^3 \quad ,
\end{eqnarray}
which gives \mbox{$-0.26283\,J$} for $J'/J=0.3$ in very good agreement with DMRG.     

\begin{figure}
\includegraphics[width=\columnwidth]{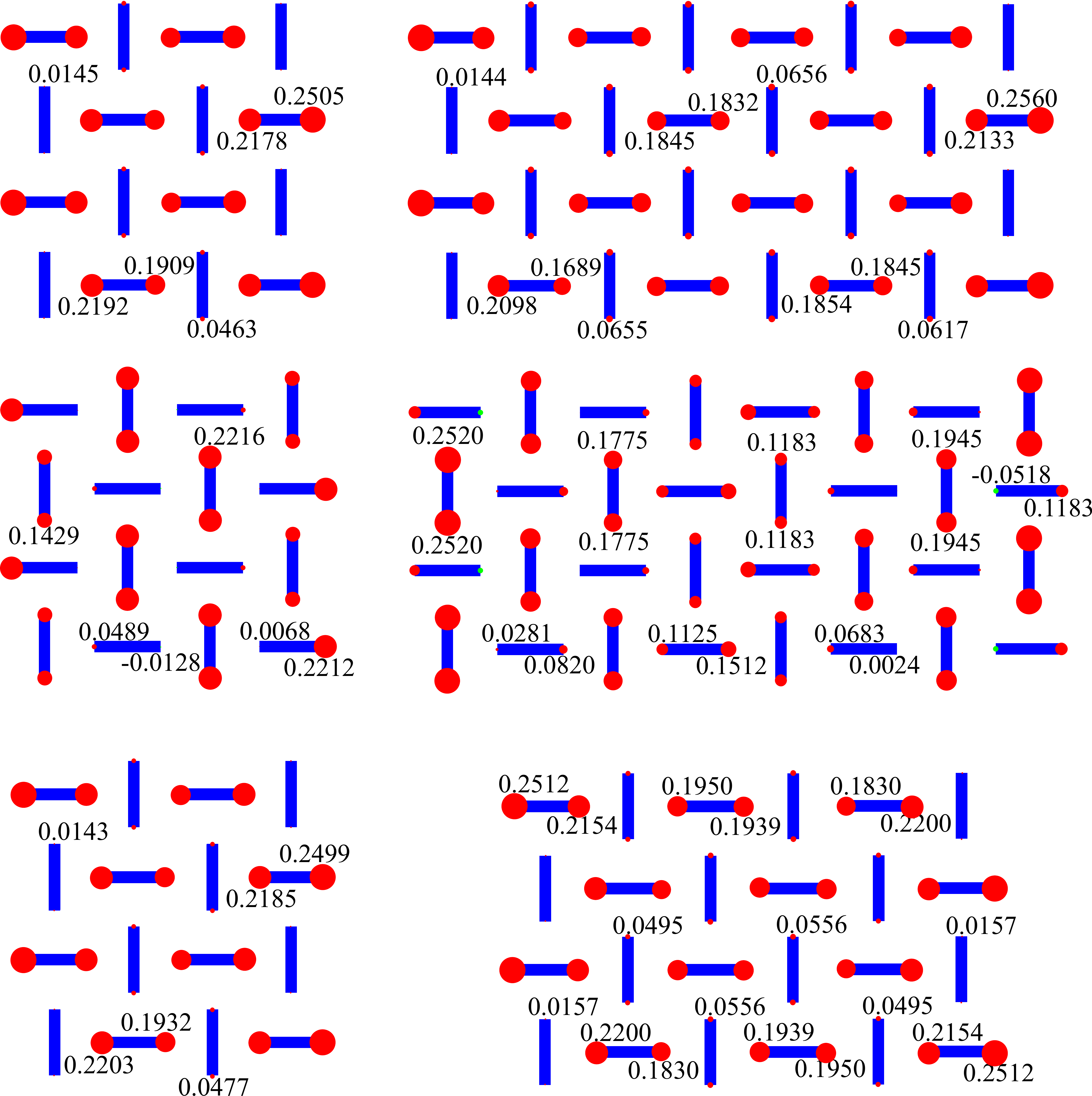}
\caption{Local magnetization of the 1/4 plateau on an open tube with 16 (left) 32 (right) dimers for $J'/J=0.3$ using DMRG (upper panel), pCUT(+ED) (middle panel), and pCUT$_{\rm finite}$(+ED) (lower panel). The diameters of circles relate to the local magnetization which is also given explicitly as numbers. Positive (negative) values denote a local magnetization parallel (antiparallel) to the field direction.}
\label{fig:0.25_pCUT_ED}
\end{figure}

Up to now we have considered the symmetry-broken state in the thermodynamic limit. On a finite open cluster, there are four equivalent diagonal stripes and the ground state is a symmetric superposition of the four states. As a consequence, the densities on all dimers of the stripes have to be the same. This is exactly what we find using the pCUT(+ED) approach on clusters with 16 and 32 dimers as displayed in the middle panel of Fig.~\ref{fig:0.25_pCUT_ED} for $J'/J=0.3$.

As mentioned above, DMRG finds a similar magnetization profile up to the fact that the role of vertical and horizontal dimers is interchanged (see also upper panel of Fig.~\ref{fig:0.25_pCUT_ED}). Exactly the same kind of pattern is also deduced with pCUT$_{\rm finite}$(+ED) as shown in the lower panel of Fig.~\ref{fig:0.25_pCUT_ED}. The reason of this discrepancy between the finite-size calculations of the full Shastry-Sutherland-Model (DMRG/pCUT$_{\rm finite}$(+ED)) and the finite-size ED of the effective model derived by pCUT in the thermodynamic limit lies in the different chemical potentials as discussed already in the method section. 

%
%%%%%%%%%%%%%%
%%%%%%%%%%%%%%
% 

In the pCUT(+ED), the chemical potential of a particle on a vertical dimer is slightly lower than the one for a particle on a horizontal dimer. Therefore, particles are placed on vertical dimers. In contrast, although the same effect is present for pCUT$_{\rm finite}$(+ED) in the bulk of the finite cluster, the chemical potential for particles on horizontal dimers is lower on the edge of the cluster compared to the corresponding one on vertical dimers. Since this energy difference is very large, it is natural to place particles on horizontal dimers when treating the four-leg Shastry-Sutherland tube on finite clusters.   
%
%%%%%%%%%%%%%%
%%%%%%%%%%%%%%
% 

Altogether, we find very strong evidence for a \mbox{$M=1/4$} magnetization plateau which corresponds to a semi-classical diagonal stripe winding around the tube. As for the plateau at \mbox{$M=1/8$}, correlated hopping is essential to stabilize this structure.
% 
%
%%%%%%%%%%%%%%
\subsection{Intermediate regime}
%%%%%%%%%%%%%%
%
%
The magnetization plateaus at $M=1/8$ and $M=1/4$ are robust features for the four-leg Shastry-Sutherland tube and we were able to find a consistent description between DMRG, pCUT$_{\rm finite}$(+ED), and pCUT(+ED). One may expect additional structures in between these two plateaus as well as between $1/4$ and $1/2$. 

In the latter region, we find some signatures for plateaus, but these are either difficult to pin down or resemble classical structures. Here, we prefer therefore to focus onto the region $M<1/4$ and leave this region of the magnetization curve for future studies.

For the regime $1/8<M<1/4$ it is also very difficult for the techniques applied by us to fully resolve the magnetization curve. Nevertheless, we would like to argue i) against the stability of three-particle wheels but ii) in favor of a plateau at $M=3/16$ which again benefits from correlated hopping. 

\begin{figure}
\includegraphics[width=0.97\columnwidth]{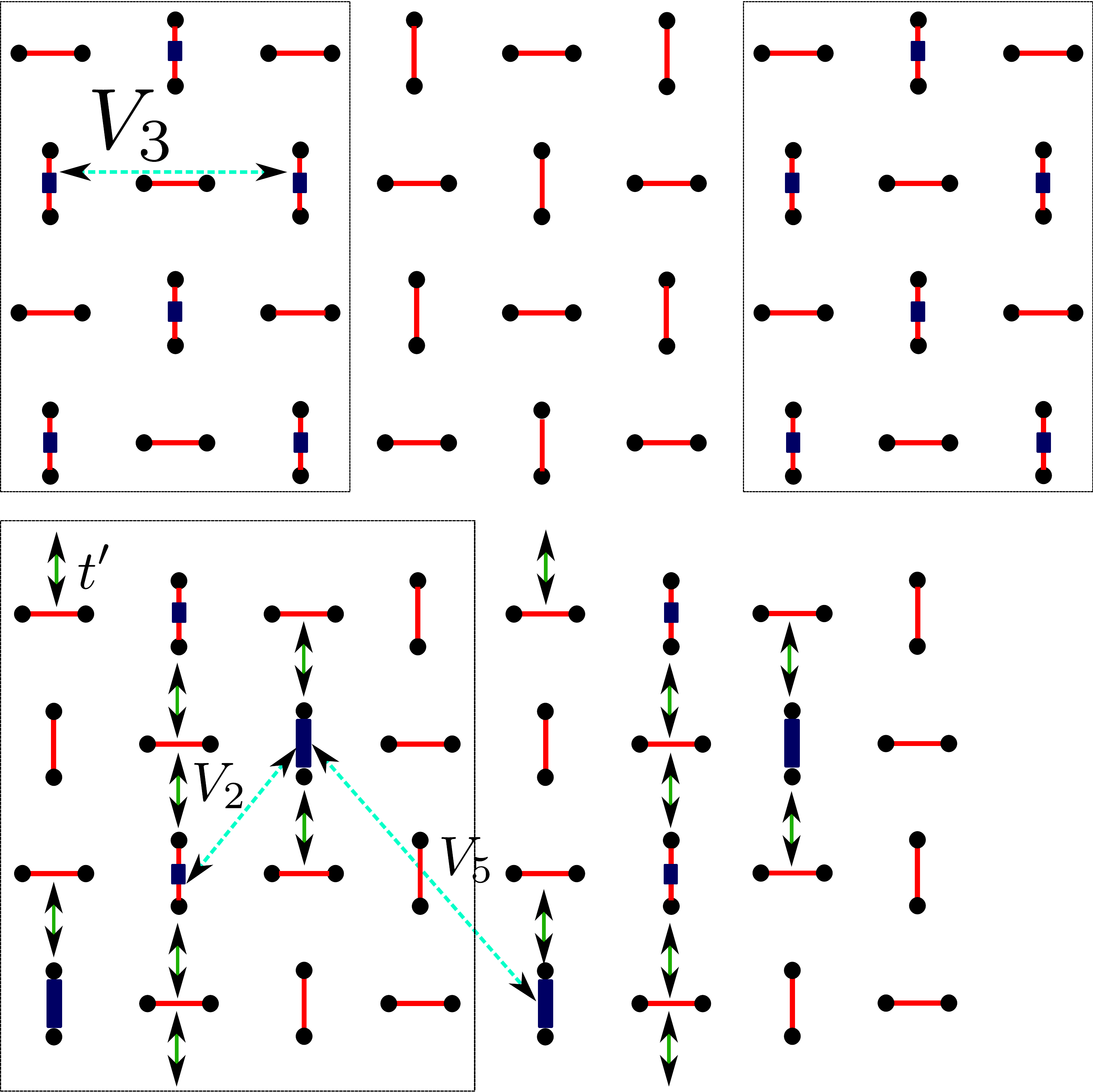}
\caption{The upper panel illustrates a crystal of three-particle wheels having $M=1/8$. Wheels are highlighted by black boxes. The lower panel shows a plateau with $M=3/16$ consisting of a classical $M=1/8$ diamond structure which is doped by a third particle in each unit cell which delocalize around the tube by correlated hopping. Thin red (thick dark) lines correspond to empty (filled) dimers. Delocalized particles are sketched as thick dark lines of half length.}
\label{fig:0.1875_SemiCA}
\end{figure}

Since wheels of the 1/8 plateau contain 2 particles and the diagonal stripe of the 1/4 plateau has 4 particles in a unit cell, one might wonder whether one can construct a magnetization profile with three particles in a unit cell. The first idea is to create the analogue of a wheel, i.e.~a highly entangled superposition of states where three particles build stripes in three subsequent rows as sketched in the upper panel of Fig.~\ref{fig:0.1875_SemiCA}. Indeed, there are several low-energy configurations of three particles on vertical dimers so that i) only interactions $V_2$ have to be paid and ii) the dominant order-two correlated hopping can act. If a crystal of such three-particle wheels should be realized in the thermodynamic limit, then the interactions between neighboring wheels must be very small. This implies three empty rows in order to avoid paying the dominant repulsions $V_1$ and $V_3$. As a consequence, the interaction between wheels is essentially zero. One can then use pCUT(+ED) on a single 12-dimer cluster to get the energy per spin $-0.31912\, J$ for $J'/J=0.3$ in the thermodynamic limit for this structure with $M=1/8$. This crystal of three-particle wheels is therefore not realized in the four-leg Shastry-Sutherland tube, since its energy is larger than the corresponding one of two-particle wheels discussed above. Physically, the relatively large energy of three-particle wheels compared to two-particle wheels comes mainly from larger energy costs due to strong repulsive density-density interactions between the three particles inside the entangled wheel. Indeed, there is a large particle density on nearest-neighbor vertical dimers such that one has to pay the large interaction $V_3$.   
   
%
%%%%%%%%%%%%%%
%%%%%%%%%%%%%%
% 

Next we discuss a crystal with magnetization \mbox{$M=3/16$} which is a promising candidate for the intermediate regime between 1/8 and 1/4, at least for not too large values of $J'/J$. The corresponding structure benefits again from correlated hopping, but at the same time it displays features of a classical plateau (see lower panel of Fig.~\ref{fig:0.1875_SemiCA}). One builds a classical $M=1/8$ structure having a unit cell of 16 dimers so that one pays only $V_5$ interactions. This classical structure is the analogue of the $M=1/8$ plateau with a diamond unit cell discussed for the 2D Shastry-Sutherland model \cite{Dorier08}. Now the third particle is doped inside each unit cell so that it can fully delocalize on two vertical dimers of the same row resulting in a structure with $M=3/16$. For small values of $J'/J$, there are i) very small energy costs between classical particles having no doped particles in between, ii) the doped particle completely delocalizes around the tube due to correlated hopping building a single-particle wheel, and iii) each classical particle has one fluctuation channel benefitting from the dominant order-two correlated hopping process as illustrated in the lower panel of Fig.~\ref{fig:0.1875_SemiCA}.   

One can estimate the energy per spin of such a $3/16$ plateau for small values of $J'/J$. We therefore take the two low-energy states with an energy $E^{\rm l}_{3/16}=3\mu+2V_2$ where three particles are located in diagonal stripes oriented from left down to right up in the lower part of the tube or from left up to right down in the upper part of the tube. Furthermore, there are six high-energy states. Two states (four states) with an energy $E^{\rm h1}_{3/16}=3\mu+V_1+V_5$ ($E^{\rm h2}_{3/16}=3\mu+V_1+V_2+V_4$) where the inner (outer) particle(s) hops due to the correlated hopping process $t'$. In total, one finds the finite matrix 
\begin{align}
 \begin{pmatrix}
 E^{\rm l}_{3/16} & 0 & t' & t' & + t' & + t' & 0 & 0 \\
 0 & E^{\rm l}_{3/16}  & t' & t' & 0 & 0 & + t' & + t' \\
 t' & t' & E^{\rm h1}_{3/16}  & 0 & 0 & 0 & 0 & 0 \\
 t' & t' & 0 & E^{\rm h1}_{3/16}  & 0 & 0 & 0 & 0 \\
 + t' & 0 & 0 & 0 & E^{\rm h2}_{3/16}  & 0 & 0 & 0 \\
 + t' & 0 & 0 & 0 & 0 & E^{\rm h2}_{3/16}   & 0 & 0 \\
 0 & + t' & 0 & 0 & 0 & 0 & E^{\rm h2}_{3/16}   & 0 \\
 0 & + t' & 0 & 0 & 0 & 0 & 0 & E^{\rm h2}_{3/16}   \\
 \end{pmatrix}% \\ &= 
\end{align}
and one obtains the ground-state energy per spin perturbatively as
\begin{eqnarray}
\frac{\epsilon_{0}^{\rm 3/16}}{J} &\approx & -\frac{9}{32}-\frac{3}{32}\left(\frac{J'}{J}\right)^2-\frac{7}{128}\left(\frac{J'}{J}\right)^3 \quad .
\end{eqnarray}
As for the other plateaus with $M=1/8$ and $M=1/4$ we observe strong binding effects due to correlated hopping. Setting $J'/J=0.3$, one obtains $\epsilon_{0}^{\rm 3/16}=-0.29116\,J$.

Let us investigate whether this plateau with $M=3/16$ is realized in the magnetization curve by comparing it to the plateaus at $M=1/8$ and $M=1/4$. In the limit of small $J'/J$, this can be done via the order-three series of the ground-state energy per spin which we have given for all three plateaus. If one defines the magnetizations $M_n(h)=\epsilon_{0}^{n}-nh$ with $n\in\{1/8,3/16,1/4\}$ as a function of the magnetic field $h$, then crossings between two structures signal first-order phase transitions. In the present case one finds that the transition between 1/8 and 3/16 takes place at \mbox{$1/2-(1/2)(J'/J)^2-(1/4)(J'/J)^3$} while the one between 3/16 and 1/4 happens at \mbox{$1/2-(1/2)(J'/J)^2-(1/16)(J'/J)^3$}. Here we have set $J=1$. Consequently, the plateau at $M=3/16$ is always realized in the limit of small $J'/J$ with a width $(3/64)(J'/J)^3$ assuming no other phase not considered here is favored.  

\begin{figure}
\includegraphics[width=0.85\columnwidth]{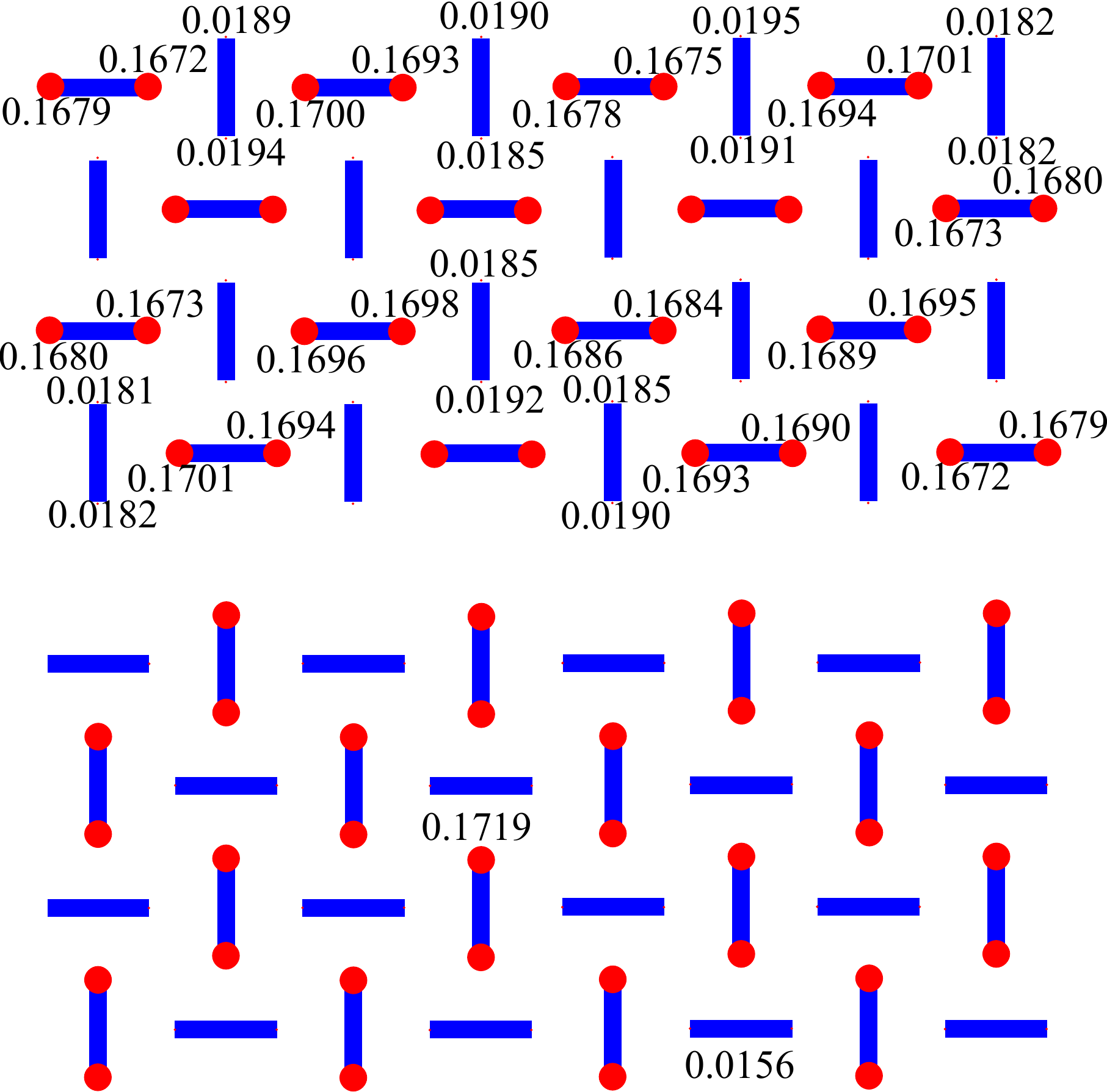}\caption{Local magnetization of the 3/16 plateau on a tube with 32 dimers for $J'/J=0.3$ using DMRG (upper panel) with periodic boundary conditions and pCUT(+ED) (lower panel). The diameters of circles relate to the local magnetization which is also given explicitly as numbers. Positive (negative) values denote a local magnetization parallel (antiparallel) to the field direction.} 
\label{fig:0.1875_pCUT_ED}
\end{figure}

Since this $M=3/16$ structure has the same 16-dimer unit cell as the two-particle wheel with $M=1/8$ and the diagonal stripe with $M=1/4$, one can also compare these plateaus on the same cluster to see which magnetization is present in the magnetization curve. The DMRG on open clusters of 32 and 64 dimers does not observe this plateau. The reason is again the very large chemical potential on vertical dimers at the edge of the clusters which makes this structure energetically disfavored. This is different for pCUT(+ED) which realizes this $M=3/16$ structure for $J'/J=0.3$ on the 32-dimer cluster with open and periodic boundary conditions (see Fig.~\ref{fig:mags} and Fig.~\ref{fig:0.1875_pCUT_ED}). One finds exactly the same ground-state energy per spin $-0.29116\,J$ for pCUT(+ED) on the periodic 32-dimer cluster strongly confirming the above considerations.  

We also performed DMRG on the same 32-dimer cluster using periodic boundary conditions. The corresponding local magnetization is shown in Fig.~\ref{fig:0.1875_pCUT_ED}. 

By applying 50 sweeps in the DMRG and keeping up to m=2500 states, we find a homogeneous solution with an energy $-0.29115\,J$ only slightly higher than the one given above. Nevertheless, the periodic boundary conditions applied here make it more difficult to converge. Indeed, we find that the values of the local magnetizations can vary by up to $3\cdot10^{-3}$ on equivalent sites, 

which is a considerably larger difference than in the presence of open boundary conditions and can be used as an error estimate for the convergence of the calculation with periodic boundary conditions. Interestingly, the structure looks very similar to the one found by pCUT(+ED) except that the role of vertical and horizontal dimers is interchanged. The DMRG result on this cluster looks like a state where effectively four particles build a classical 1/8 plateau with diamond unit cell on {\it horizontal} dimers while the remaining two particles delocalize via correlated hopping horizontally. From the considerations above, it is reasonable that such a state has also a very low energy for this cluster which is slightly above the true ground state. We expect that the energy difference between these two states increases with increasing system size, since particles delocalizing horizontally in the same row have to pay repulsive interactions. As a consequence, it might be easier for the DMRG to resolve this structure, in which the vertical dimers are polarized, on larger systems.   
%
%%%%%%%%%%%%%%
%%%%%%%%%%%%%%
% 

Interestingly, the plateau at $M=3/16$ is not present within pCUT(+ED) for a larger ratio $J'/J=0.5$. Physically, we suspect that this structure, which combines elements of a classical plateau and a highly entangled one-particle wheel, does not benefit as much from correlated hopping with increasing $J'/J$ as the two-particle wheel with $M=1/8$ (see Fig.~\ref{fig:mags}).   
% 
%
%%%%%%%%%%%%%%
\section{Implications for 2D}
%%%%%%%%%%%%%%
%
%
In this section we discuss how our findings for the quasi-2D Shastry-Sutherland 
tube relate to the physics of the two-dimensional Shastry-Sutherland model which is 
believed to be a good microscopic model for the frustrated quantum magnet SrCu(BO$_3$)$_2$. 
 To this end we focus on the dominant plateaus at $M=1/8$ and $M=1/4$, since the plateau at 
$M=3/16$ is only present for small $J'/J$ in the tube while the origin of the $M=1/2$ 
is anyway not debated.

\subsection{$M=1/8$}
Let us start with the interesting plateau at $M=1/8$ of the four-leg Shastry-Sutherland tube 
consisting of highly entangled wheels oriented transverse to the tube direction. Each 16-dimer unit cell contains one wheel where two particles are essentially delocalized over the inner four vertical dimers. 

This plateau is certainly specific to the tube geometry. Indeed, due to the finite transverse extension a {\it finite} number of triplons in one wheel is sufficient to construct a structure at {\it finite density}. It becomes then preferable for the four-leg tube to fully delocalize the two triplons in such a wheel in order to benefit maximally from correlated hopping. 

Now imagine we increase the number of legs $N_{\rm legs}$ of the tube up to infinity which corresponds to the two-dimensional case. If one creates the same kind of state as above, i.e.~we take a rectangular unit cell of size $4\times N_{\rm legs}$ which contains two delocalized triplons, one obtains a magnetization $2/(4N_{\rm legs})$ which decreases gradually with $N_{\rm legs}$. For $N_{\rm legs}=6$, one then expects a similar plateau at $M=1/12$, for $N_{\rm legs}=8$ one has a plateau at $M=1/16$, and so on. But in the limit $N_{\rm legs}\rightarrow\infty$ this
 kind of state corresponds to a zero-density state and is thus irrelevant for the two-dimensional system.

\begin{figure}
\includegraphics[width=\columnwidth]{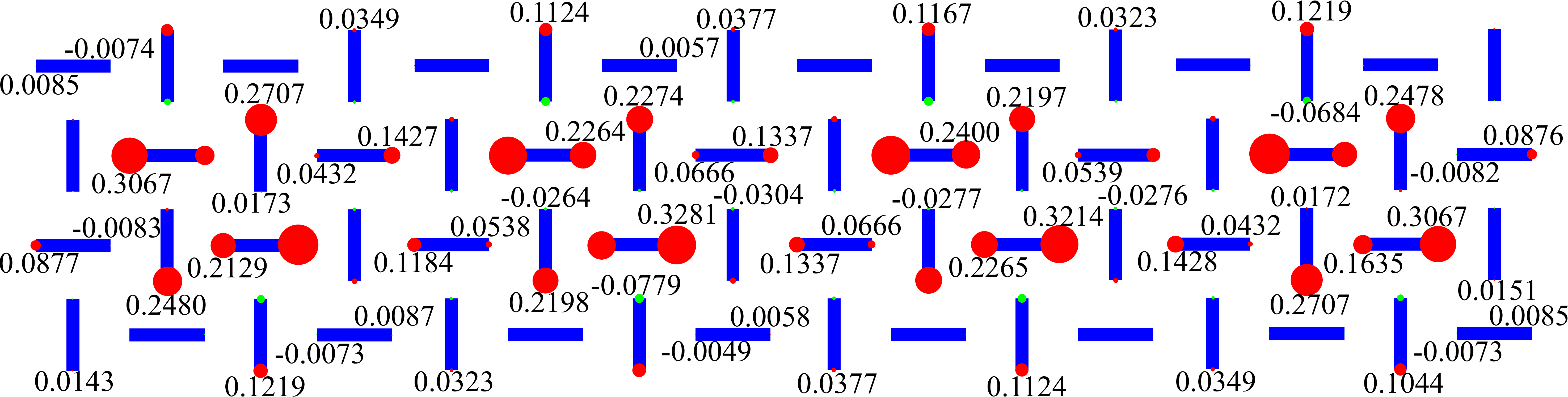}
\caption{Local magnetization of the 1/8 plateau on a four-leg system with 64 dimers for $J'/J=0.3$ using open boundary conditions in both directions obtained by DMRG. The diameters of circles relate to the local magnetization which is also given explicitly as numbers. Positive (negative) values denote a local magnetization parallel (antiparallel) to the field direction.} 
\label{fig:0.125_DMRG_OBC}
\end{figure}

Nevertheless, we would like to stress that although the wheels are specific to the tube geometry, the mechanism, which stabilizes the wheels, is very likely relevant for the 2D Shastry-Sutherland model. Indeed, the recently discovered crystals of bound states for the 2D problem \cite{Corboz14} have exactly the same quantum numbers as our crystal of wheels, i.e.~in both systems two-particle bound states with $S^{z}_{\rm tot}=2$ are found. It is therefore tempting to interpret the pinwheel structures in 2D as two particles gaining kinetic energy due to correlated hopping processes. Furthermore, we observe similar pinwheel structures also in the four-leg system when applying open boundary conditions to the both direction (see Fig.~\ref{fig:0.125_DMRG_OBC}). While we leave a detailed study of this case to future studies, this finding again indicates that our physical picture obtained for the four-leg Shastry-Surtherland tube is likely of relevance for the full 2D case.   

%
%%%%%%%%%%%%%%
%%%%%%%%%%%%%%
% 

\subsection{$M=1/4$}

For $M=1/4$, we find that correlated hopping stabilizes a semiclassical structure consisting of a diagonal stripe which wraps around the tube. This structure does not correspond to the classical solution. Quantum fluctuations induced by correlated hopping are essential to lower the energy of this structure.

Interestingly, such diagonal stripes have been proposed for the $M=1/4$ plateau of the two-dimensional model as well as for the experimental findings for the compound SrCu(BO$_3$)$_2$ \cite{Miyahara03b,Takigawa13,Matsuda13,Corboz14}, but no microscopic mechanism was yet deduced for this structure, e.g.~no plateau at $M=1/4$ has been found in Ref.~\onlinecite{Dorier08}, where the classical limit of the effective two-dimensional model has been investigated in detail. Our microscopic results for the four-leg Shastry-Sutherland tube indicate that this discrepancy is due to neglecting quantum fluctuations induced by correlated hopping. Furthermore, we expect that the energy per site of the $M=1/4$ plateau on the tube should be very close to the one of the corresponding two-dimensional structure. The reason is that the two-dimensional structure fits perfectly on the four-leg tube geometry and virtual fluctuations feeling the finite transverse extension of the tube are clearly of subleading order.

% 
%
%%%%%%%%%%%%%%
\section{Conclusions}
%%%%%%%%%%%%%%
%
%
In this work we have studied the magnetization process of a four-leg Shastry-Sutherland
 tube by DMRG, pCUT(+CA/+ED), and pCUT$_{\rm finite}$(+ED). Most importantly, we have identified unconventional magnetization plateaus at \mbox{$M=1/8$} and \mbox{$M=1/4$} which do not correspond to classical structures of frozen triplons. In all cases quantum fluctuations induced by correlated hopping processes of triplons are essential. We stress that both plateaus are stable in a broad range of parameters.

The nature of the two plateaus is strikingly different. The plateau at $M=1/4$ is understood semi-classically by a diagonal stripe of triplons wrapping around the tube. It is this structure which benefits most from correlated hopping. In contrast, the low-magnetization plateau at $M=1/8$ consists of highly entangled transverse wheels where each wheel contains a two-triplon bound state which fully delocalizes around the tube. 

The intermediate regimes between $M=1/8$ and \mbox{$M=1/4$} as well as the one between $M=1/4$ and \mbox{$M=1/2$} are very demanding to pinpoint. Here one is facing the problem that different structures with already large unit cells may build super-structures with even larger unit cells which are very hard to treat by any theoretical method. We think it might be an interesting option to use DMRG or other variational tools directly on the effective model derived by pCUT in order to tackle this question at least in the limit of small $J'/J$ where the effective model is fully converged. At the same time this may shed light on the appearance of pair superfluids at very low magnetizations or supersolid phases in the intermediate regimes. Indeed, correlated hopping is known to be able to stabilize such phases \cite{Momoi00b,Bendjama05,Schmidt06,Schmidt08}.  

Our work shows that correlated hopping processes, or more generally quantum fluctuations, are important also for the two-dimensional case. This aspect is very likely the reason between the discrepancies of the pCUT(+CA) approach used in Ref.~\onlinecite{Dorier08} finding magnetization plateaus of single triplons and the recently discovered sequence of magnetization plateaus of two-particle bound states \cite{Corboz14}. It is therefore not the effective model derived by pCUT which fails for the crystal of bound states, but it is clearly the CA which is not able to treat correlated hopping processes well as we have seen impressively for the Shastry-Sutherland four-leg tube.  

Altogether, our study of the four-leg Shastry-Sutherland tube reveals that the character of magnetization plateaus can indeed be given by delocalized structures of multiple triplons which are stabilized by exotic quantum fluctuations, as e.g. correlated hopping. It will be intriguing to search for such effects in further systems. 

%%%%%%%%%%%%%%%%%%%%%%%%%%%%%%%%%%%%%%%%%%%%%%%%%%%%%%%%%%%%%%%%%%%%%%%%%%%%%%%%%%%%%%%%%%%%%%%%%%%%%%%%%%%%%%%%%%%%%
\section{Acknowledgements}
%%%%%%%%%%%%%%%%%%%%%%%%%%%%%%%%%%%%%%%%%%%%%%%%%%%%%%%%%%%%%%%%%%%%%%%%%%%%%%%%%%%%%%%%%%%%%%%%%%%%%%%%%%%%%%%%%%%%%
We acknowledge very useful discussions with F.~Mila and S.~Wessel. We also thank S.~Wessel for providing us QMC data for related sign-free systems which helped optimizing the DMRG procedure. KPS acknowledges ESF and EuroHorcs for funding through his EURYI. \newpage
\onecolumngrid

%%%%%%%%%%%%%%%%%%%%%%%%%%%%%%%%%%%%%%%%%%%%%%%%%%%%%%%%%%%%%%%%%%%%%%%%%%%%%%%%%%%%%%%%%%%%%%%%%%%%%%%%%%%%%%%%%%%%%
\section{Appendix}
%%%%%%%%%%%%%%%%%%%%%%%%%%%%%%%%%%%%%%%%%%%%%%%%%%%%%%%%%%%%%%%%%%%%%%%%%%%%%%%%%%%%%%%%%%%%%%%%%%%%%%%%%%%%%%%%%%%%%
\label{sect:Appendix}

%\subsection{Classical plateau energies}
%In the following we list the explicit expressions for the classical energy
% per dimer $\epsilon_{\rm cl}=E_{\rm cl}/(JN)$ of all considered plateaus:
%\begin{eqnarray*}
% \epsilon^{1/6}_{\rm cl, ca} &=& \frac{1}{6} \left( V^\prime_3 + 2V_7\right) \\
% \epsilon^{1/6}_{\rm cl,square} &=& \frac{1}{6} \left( V_4+V_5 +V_6\right) \\
% \epsilon^{1/6}_{\rm cl,stripe} &=& \frac{1}{6} \left( V_4+V_5 +V_6\right) \\
% \epsilon^{1/6}_{\rm cl,new} &=& \frac{1}{6} \left( V_4+V_5 +V_6\right) \\
% \epsilon^{2/15}_{\rm cl,rect} &=& \frac{1}{15} \left( V^\prime_3+V_7 +2V_6 + 2V_8\right) \\
% \epsilon^{2/15}_{\rm cl,rhomb} &=& \frac{1}{15} \left( V^\prime_3+V_7 +2V_6 + 2V_8\right) \\
%\epsilon^{2/15}_{\rm cl, big} &=& \frac{1}{30} \left( V_4+4V_5+ V_6 + 3V^\prime_{7} + 3V_7 \right) \\
% \epsilon^{2/15}_{\rm cl, b_2} &=& \frac{1}{30} \left( 7V_5 + V_4 + V_6 \right) 
%\end{eqnarray*}
%\begin{eqnarray*}
%\epsilon^{1/8}_{\rm cl,square} &=& \frac{1}{4} V_5 \\
% \epsilon^{1/8}_{\rm cl,rhomboid} &=& \frac{1}{8} \left( V_5+V_7 \right) \\
% \epsilon^{1/8}_{\rm cl,ca} &=& \frac{1}{24} \left( V^\prime_{3} + 4V_6 + V7 + 2V_8 \right) \\
% \epsilon^{1/9}_{\rm cl} &=& \frac{1}{9} 2V_6 \quad .
%\end{eqnarray*}
In this appendix we give the series expansions of the most relevant amplitudes of the effective pCUT model. One-body operators (two-body operators) are calculated up to order 15 (14) in the parameter $x\equiv J'/J$. Note that we have set $J=1$ in all expressions.  
\subsection{One-body operators}
In the following we show the series expansions for the chemical potential on vertical and horizontal dimers as well as the hopping over the diagonal:
\begin{align*}
t_{(0,0)}^{\rm v} &=1- x ^2 - \frac{1}{2}  x ^3- \frac{1}{8}  x ^4+ \frac{5}{32}  x ^5+ \frac{3}{128}  x ^6- \frac{1699}{4608}  x ^7- \frac{35107}{55296}  x ^8- \frac{259061}{663552}  x ^9 + \frac{974687}{6635520}  x ^{10} + \frac{1151870527}{4777574400} x ^{11} \notag \\ &\quad -\frac{23323367629}{38220595200}  x ^{12} - \frac{40392669400271}{22932357120000}  x ^{13}- \frac{102289876433461163}{57789539942400000}  x ^{14} - \frac{8204339820020446111}{48543213551616000000}  x ^{15} \\
t_{(0,0)}^{\rm h} &=1- x ^2 - \frac{1}{2}  x ^3- \frac{1}{8}  x ^4+ \frac{5}{32}  x ^5+ \frac{3}{128}  x ^6- \frac{1699}{4608}  x ^7- \frac{35107}{55296}  x ^8- \frac{259061}{663552}  x ^9 + \frac{194755}{1327104}  x ^{10} + \frac{1153266463}{4777574400} x ^{11} \notag \\ &\quad +\frac{116030601329}{191102976000}  x ^{12} + \frac{40140647360519}{22932357120000}  x ^{13}+ \frac{20172410628174247}{11557907988480000}  x ^{14} + \frac{12277459855094982833}{59708642710323200000}  x ^{15}\\
t_{(1,1)}^{\rm v} &= -\frac{1}{96}  x ^6- \frac{11}{576} x ^7 - \frac{83}{4608}  x ^8- \frac{2447}{663552}  x ^9- \frac{10487557}{79626240}  x ^{10} - \frac{303150173}{9555148800}  x ^{11} -\frac{6754465217}{76441190400}  x ^{12} \notag \\ &\quad - \frac{14930736446759}{137594142720000}  x ^{13} - \frac{212436546502069}{4280706662400000}  x ^{14} + \frac{1767219308670276631}{97086427103232000000}  x ^{15} \\
t_{(1,1)}^{\rm h} &= -\frac{1}{96}  x ^6- \frac{11}{576} x ^7 - \frac{83}{4608}  x ^8- \frac{2447}{663552}  x ^9- \frac{10487557}{79626240}  x ^{10} - \frac{303150173}{9555148800}  x ^{11} -\frac{6759171137}{76441190400}  x ^{12} \notag \\ &\quad - \frac{149631502102793}{137594142720000}  x ^{13} - \frac{1133557569961}{22649241600000}  x ^{14} + \frac{1750239587860726951}{97086427103232000000}  x ^{15}
\end{align*}
with $t_{(1,1)}^{\rm v}=t_{(-1,1)}^{\rm v}$ and $t_{(1,1)}^{\rm h}=t_{(-1,1)}^{\rm h}$. In the main body of the paper we often use $\mu$ for the chemical potenzial on all dimers which reflects the fact that $t_{(0,0)}^{\rm v}$ and $t_{(0,0)}^{\rm h}$ only differ in order 10.
\subsection{Two-body density-density interactions}
The type of density-density interactions called $V_1$ in the main body of the text is given by:
\begin{align*}
V_{(1,0),(0,0),(1,0)}^{\rm v} &=  \frac{1}{2}  x + \frac{1}{2}  x ^2- \frac{1}{8}  x ^3- \frac{9}{16}  x ^4- \frac{3}{64}  x ^5+ \frac{809}{768}  x ^6+ \frac{2173}{3072}  x ^7- \frac{70543}{24576}  x ^8- \frac{37816411}{5308416}  x ^9 \notag \\
&\quad - \frac{2055058321}{637009920}  x ^{10} + \frac{12212246377}{76441190400}  x ^{11} +\frac{329845478498011}{9172942848000}  x ^{12} \notag \\ &\quad + \frac{52342527662776237}{1100753141760000}  x ^{13}+ \frac{19409208366246467731}{924632639078400000}  x ^{14} \\
V_{(0,1),(0,0),(0,1)}^{\rm v} &=  \frac{1}{2}  x + \frac{1}{2}  x ^2- \frac{1}{8}  x ^3- \frac{9}{16}  x ^4- \frac{3}{64}  x ^5+ \frac{809}{768}  x ^6+ \frac{2173}{3072}  x ^7- \frac{211309}{73728}  x ^8- \frac{37640899}{5308416}  x ^9 \notag \\ 
&\quad - \frac{396983309}{127401984}  x ^{10} + \frac{341825781751}{25480396800}  x ^{11} +\frac{62793791370601}{3057647616000}  x ^{12} \notag \\ &\quad + \frac{544988441572090729}{1100753141760000}  x ^{13}+ \frac{30729593506190988097}{924632639078400000}  x ^{14}\quad .
\end{align*}
The type of density-density interactions called $V_3$ in the main body of the text read:
\begin{align*}
V_{(2,0),(0,0),(2,0)}^{\rm v} &=  \frac{1}{2}  x ^2+ \frac{3}{4}  x ^3- \frac{1}{8}  x ^4- \frac{49}{64}  x ^5- \frac{289}{768}  x ^6+ \frac{4019}{9216}  x ^7+ \frac{77609}{110592}  x ^8+ \frac{243991}{1327104}  x ^9 \notag \\
&\quad - \frac{73855279}{79626240}  x ^{10} - \frac{1584489421}{1061683200}  x ^{11}- \frac{1756843}{119439360}  x ^{12} \notag \\ &\quad + \frac{100261859756351}{45864714240000}  x ^{13}+ \frac{183832105244723489}{115579079884800000}  x ^{14} \\
V_{(0,2),(0,0),(0,2)}^{\rm h} &=  \frac{1}{2}  x ^2+ \frac{3}{4}  x ^3- \frac{5}{16}  x ^4- \frac{11}{64}  x ^5+ \frac{73}{768}  x ^6- \frac{1729}{3072}  x ^7+ \frac{47567}{36864}  x ^8- \frac{386201}{1327104}  x ^9 \notag \\
&\quad- \frac{157817}{81920}  x ^{10} + \frac{25294352587}{9555148800}  x ^{11} -\frac{5584694533789}{1146617856000}  x ^{12} \notag \\ &\quad + \frac{46040943841567}{27518828544000}  x ^{13}+ \frac{155882295114023201}{19263179980800000}  x ^{14} \quad .
\end{align*}
The type of density-density interactions called $V_2$ in the main body of the text read:
\begin{align*}
V_{(1,1),(0,0),(1,1)}^{\rm v} &=  \frac{1}{4}  x ^3+ \frac{3}{8}  x ^4+ \frac{23}{64}  x ^5- \frac{41}{128}  x ^6- \frac{337}{192}  x ^7- \frac{283327}{221184}  x ^8+ \frac{23684687}{5308416}  x ^9 \notag \\
&\quad + \frac{1362906325}{127401984}  x ^{10} - \frac{12212246377}{76441190400}  x ^{11} -\frac{329845478498011}{9172942848000}  x ^{12} \notag \\ &\quad - \frac{52342527662776237}{1100753141760000}  x ^{13}+ \frac{19409208366246467731}{924632639078400000}  x ^{14} \\
V_{(1,1),(0,0),(1,1)}^{\rm h} &=  \frac{1}{4}  x ^3+ \frac{3}{8}  x ^4+ \frac{23}{64}  x ^5- \frac{39}{128}  x ^6- \frac{971}{576}  x ^7- \frac{262199}{221184}  x ^8+ \frac{23806187}{5308416}  x ^9 \notag \\
&\quad + \frac{6742797373}{637009920}  x ^{10} - \frac{15167380493}{76441190400}  x ^{11} -\frac{326101716505943}{9172942848000}  x ^{12} \notag \\ &\quad - \frac{51636671081928193}{1100753141760000}  x ^{13}+ \frac{9661058826847003663}{924632639078400000}  x ^{14} \quad .
\end{align*}
The type of density-density interactions called $V_4$ in the main body of the text read:
\begin{align*}
V_{(2,1),(0,0),(2,1)}^{\rm v} &=  \frac{1}{8}  x ^4+ \frac{17}{64}  x ^5+ \frac{77}{768}  x ^6- \frac{3571}{9216}  x ^7- \frac{59017}{110592}  x ^8+ \frac{1339739}{2654208}  x ^9+ \frac{323608631}{159252480}  x ^{10} \\
&\quad + \frac{51274985519}{38220595200}  x ^{11} -\frac{3458863679681}{1146617856000}  x ^{12}- \frac{3387691098242909}{550376570880000}  x ^{13} \notag \\ &\quad + \frac{180169037145982051}{231158159769600000}  x ^{14} \\
V_{(1,2),(0,0),(1,2)}^{\rm v} &=  \frac{1}{8}  x ^4+ \frac{17}{64}  x ^5- \frac{17}{384}  x ^6- \frac{6349}{9216}  x ^7- \frac{17633}{36864}  x ^8+ \frac{4476433}{2654208}  x ^9+ \frac{567596647}{159252480}  x ^{10} \\
&\quad - \frac{11178106469}{38220595200}  x ^{11} -\frac{18177464069}{1887436800}  x ^{12}- \frac{4776276471478103}{550376570880000}  x ^{13} \notag \\ &\quad + \frac{8711588250104481127}{462316319539200000}  x ^{14}\quad .
\end{align*}
All other density-density interactions appear in at least order 6 or higher.
\subsection{Correlated hopping}
The type of correlated hopping called $t'$ in the main body of the text is given by:
\begin{align*}
V_{(1,0),(1,0),(0,1)}^{\rm h} &=  -\frac{1}{4}  x ^2- \frac{5}{16}  x ^3+ \frac{9}{64}  x ^4+ \frac{89}{128}  x ^5+ \frac{3731}{6144}  x ^6- \frac{42367}{49152}  x ^7- \frac{9616315}{3538944}  x ^8 \notag \\
&\quad - \frac{24411431}{28311552}  x ^9 + \frac{77453067637}{10192158720}  x ^{10} + \frac{1109080209497}{81537269760}  x ^{11} \notag \\
&\quad -\frac{832683372515471}{146767085568000}  x ^{12} - \frac{102609109563591289}{1956894474240000}  x ^{13} - \frac{23343475477851433}{46965467381760000}  x ^{14} \\
V_{(0,1),(0,1),(1,0)}^{\rm v} &=  -\frac{1}{4}  x ^2- \frac{5}{16}  x ^3+ \frac{9}{64}  x ^4+ \frac{89}{128}  x ^5+ \frac{3731}{6144}  x ^6- \frac{42367}{49152}  x ^7- \frac{9647035}{3538944}  x ^8 \notag \\
&\quad - \frac{8536973}{9437184}  x ^9 + \frac{76437263413}{10192158720}  x ^{10} + \frac{611075103877}{45298483200}  x ^{11} \notag \\
&\quad -\frac{830592692720663}{146767085568000}  x ^{12} + \frac{8302601110042189391}{17612050268160000}  x ^{13}- \frac{668866209122606693}{986274815016960000}  x ^{14} \quad .
\end{align*}
The type of correlated hopping called $t''$ in the main body of the text is given by:
\begin{align*}
V_{(1,0),(1,0),(2,0)}^{\rm h} &=  \frac{1}{4}  x ^2+ \frac{3}{8}  x ^3+ \frac{1}{8}  x ^4- \frac{13}{32}  x ^5- \frac{61}{64}  x ^6- \frac{591}{1024}  x ^7+ \frac{107867}{73728}  x ^8+ \frac{6013199}{1769472}  x ^9 \notag \\
&\quad - \frac{6901591}{14155776}  x ^{10} - \frac{24515591279}{1698693120}  x ^{11} -\frac{16279526208581}{611529523200}  x ^{12} \notag \\ &\quad - \frac{96063688771843}{24461180928000}  x ^{13}+ \frac{3942846918014280953}{61642175938560000}  x ^{14} \\
V_{(0,1),(0,1),(0,2)}^{\rm v} &=  \frac{1}{4}  x ^2+ \frac{3}{8}  x ^3+ \frac{1}{8}  x ^4- \frac{13}{32}  x ^5- \frac{61}{64}  x ^6- \frac{591}{1024}  x ^7+ \frac{107227}{73728}  x ^8+ \frac{5931755}{1769472}  x ^9 \notag \\
&\quad - \frac{14923729}{23592960}  x ^{10} - \frac{374548151797}{25480396800}  x ^{11} -\frac{16400221354633}{611529523200}  x ^{12} \notag \\ &\quad - \frac{1291801992179591}{366917713920000}  x ^{13}- \frac{6066194272818874343}{154105439846400000}  x ^{14}\quad .
\end{align*}
\twocolumngrid


\begin{thebibliography}{10}
\bibitem{book_quantummagnetism}
C. Lacroix, P. Mendels, and F. Mila (Eds), {\it Introduction to Frustrated Magnetism} Springer Series in Solid-State Sciences {\bf 164}, (2011).
%\bibitem{Honecker04}  
%For an early review, see A. Honecker, J. Schulenburg, J. Richter,
%J. Phys.: Condens. Matter {\bf 16}, S749 (2004) and references therein.
\bibitem{Kageyama99}
H. Kageyama, K. Yoshimura, R. Stern, N. V. Mushnikov,
K. Onizuka, M. Kato, K. Kosuge, C. P. Slichter, T. Goto, and
Y. Ueda, Phys. Rev. Lett. {\bf 82}, 3168 (1999).
\bibitem{Onizuka00}
K. Onizuka, H. Kageyama, Y. Narumi, K. Kindo, Y. Ueda, and
T. Goto, J. Phys. Soc. Jpn. {\bf 69}, 1016 (2000).
\bibitem{Kageyama00}
H. Kageyama, M. Nishi, N. Aso, K. Onizuka, T. Yosihama,
K. Nukui, K. Kodama, K. Kakurai, and Y. Ueda, Phys. Rev.
Lett. {\bf 84}, 5876 (2000).
\bibitem{Kodama02}
K. Kodama, M. Takigawa, M. Horvatic, C. Berthier,
H. Kageyama, Y. Ueda, S. Miyahara, F. Becca, and F. Mila,
Science {\bf 298}, 395 (2002).
\bibitem{Takigawa04}
M. Takigawa, K. Kodama, M. Horvatic, C. Berthier,
H. Kageyama, Y. Ueda, S. Miyahara, F. Becca, and F. Mila,
Physica B: Condensed Matter {\bf 346}, 27 (2004).
\bibitem{Levy08}
F. Levy, I. Sheikin, C. Berthier, M. Horvatic, M. Takigawa,
H. Kageyama, T. Waki, and Y. Ueda, EPL (Europhysics Letters)
{\bf 81}, 67004 (2008).
\bibitem{Sebastian08}
S. E. Sebastian, N. Harrison, P. Sengupta, C. D. Batista, S. Francoual,
E. Palm, T. Murphy, N. Marcano, H. A. Dabkowska, and
B. D. Gaulin, Proceedings of the National Academy of Sciences
{\bf 105}, 20157 (2008).
\bibitem{Jaime12}
M. Jaime, R. Daou, S. A. Crooker, F.Weickert, A. Uchida, A. E.
Feiguin, C. D. Batista, H. A. Dabkowska, and B. D. Gaulin,
Proceedings of the National Academy of Sciences (2012),
10.1073/pnas.1200743109.
\bibitem{Takigawa13}
 M. Takigawa, M. Horvatic, T. Waki, S. Kr\"amer, C. Berthier,
F. LÂ´evy-Bertrand, I. Sheikin, H. Kageyama, Y. Ueda, and
F. Mila, Phys. Rev. Lett. {\bf 110}, 067210 (2013).
\bibitem{Matsuda13}
Y. H. Matsuda, N. Abe, S. Takeyama, H. Kageyama, P. Corboz,
A. Honecker, S. R. Manmana, G. R. Foltin, K. P. Schmidt, and
F. Mila, Phys. Rev. Lett. {\bf 111}, 137204 (2013).
\bibitem{Miyahara99}
 S. Miyahara and K. Ueda, Phys. Rev. Lett. {\bf 82}, 3701 (1999).
\bibitem{Momoi00a}
T. Momoi and K. Totsuka, Phys. Rev. B {\bf 61}, 3231 (2000).
\bibitem{Momoi00b}
T. Momoi and K. Totsuka, Phys. Rev. B 62, 15067 (2000).
\bibitem{Fukumoto00}
Y. Fukumoto and A. Oguchi, J. Phys. Soc. Jpn. {\bf 69}, 1286 (2000).
\bibitem{Fukumoto01}
Y. Fukumoto, J. Phys. Soc. Jpn. {\bf 70}, 1397 (2001).
\bibitem{Miyahara03a}
S. Miyahara and K. Ueda, J. Phys.: Condensed Matter {\bf 15}, R327
(2003).
\bibitem{Miyahara03b}
S. Miyahara, F. Becca, and F. Mila, Phys. Rev. B {\bf 68}, 024401
(2003).
\bibitem{Dorier08}
J. Dorier, K.P. Schmidt and F. Mila, Phys. Rev. Lett. {\bf 101}, 250402 (2008).
\bibitem{Abendschein08}
A. Abendschein and S. Capponi, Phys. Rev. Lett. {\bf 101}, 227201 (2008).
\bibitem{Nemec12}
M. Nemec, G. R. Foltin, and K. P. Schmidt, Phys. Rev. B {\bf 86},
174425 (2012).
\bibitem{Lou12}
J. Lou, T. Suzuki, K. Harada, and N. Kawashima, arXiv:1212.1999 (2012).
\bibitem{Corboz14}
P. Corboz and F. Mila, Phys. Rev. Lett. {\bf 112}, 147203 (2014).
\bibitem{Shastry81}
B.S. Shastry and B. Sutherland, Physica (Amsterdam) {\bf 108} B+C, 1069 (1981).
\bibitem{Knetter00}
C. Knetter and G. S. Uhrig, Eur. Phys. J. B {\bf 13}, 209 (2000).
\bibitem{Knetter03}
C. Knetter, K.P. Schmidt, and G.S. Uhrig, J. Phys. A: Math. and Gen. {\bf 36}, 7889 (2003).
\bibitem{DMRG}
S.R. White, Phys. Rev. Lett. {\bf 69}, 2863 (1992); U. Schollw\"ock, Rev. Mod. Phys. {\bf 77}, 259 (2005). 
\bibitem{WhiteScienceKagome}
S. Yang, D.A. Huse, and S.R. White, Science {\bf  332}, 1173 (2011). 
\bibitem{Stoudenmire2D}
E.M. Stoudenmire and S.R. White,  Annu. Rev. Condens. Matter Phys. 2012. 3:111Ð28 . 
\bibitem{Manmana11}
S.R. Manmana, J.-D. Picon, K.P. Schmidt, and F. Mila, Eur. Phys. Lett. {\bf 94}, 67004 (2011).
\bibitem{Schmidt03}
K.P. Schmidt and G.S. Uhrig, Phys. Rev. Lett. {\bf 90}, 227204 (2003).
\bibitem{Matsubara56}
T. Matsubara and H. Matsuda, Prog. Theor. Phys. 16, 569 (1956). 
\bibitem{Lanczos50}
K. Lanczos, J. Res. Natl. Bur. Stand {\bf 45}, 225 (1950); J. K. Cullum and R. A. Willoughby, {\it Lanczos Algorithms for Large Symmetric Eigenvalue Computations (SIAM}, Philadelphia, (2002).
\bibitem{WhitePRL2007}
S.R. White and A.L. Chernyshev, Phys. Rev. Lett. {\bf 99}, 127004 (2007).
\bibitem{Corboz13}
P. Corboz and F. Mila, Phys. Rev. B  {\bf 87}, 115144 (2013). 
\bibitem{Bendjama05}
R. Bendjama, B. Kumar, and F. Mila, Phys. Rev. Lett. {\bf 95}, 110406 (2005).
\bibitem{Schmidt06}
K.P. Schmidt, J. Dorier, A. L\"auchli, and F. Mila, Phys. Rev. B {\bf 74}, 174508 (2006).
\bibitem{Schmidt08}
K.P. Schmidt, J. Dorier, A. L\"auchli and F. Mila, Phys. Rev. Lett. {\bf 100}, 090401 (2008).
\end{thebibliography}
\end{document}